\documentclass[5p,times]{elsarticle}
\usepackage{subfig}
\usepackage{adjustbox}
\usepackage{url}
\usepackage[utf8]{inputenc}
\usepackage{amssymb}
\usepackage{amsmath}
\usepackage[disable]{todonotes}
\usepackage{makecell}
\usepackage{tabularx}
\usepackage{algorithm}
\usepackage{algpseudocode}
\usepackage{changes}
\begin{document}

\title{Cost-efficient Auto-scaling of Container-based Elastic Processes}

\author[1]{Gerta Sheganaku}
\author[2]{Stefan Schulte}
\ead{stefan.schulte@tuhh.de}
\author[1]{Philipp Waibel}
\author[4]{Ingo Weber}
\address[1]{TU Wien, Vienna, Austria}
\address[2]{Christian Doppler Laboratory for Blockchain Technologies for the Internet of Things, Hamburg University of Technology, Hamburg, Germany}

%\affiliation[3]{organization={WU Wien}, city={Vienna}, country={Austria}}
\address[4]{TU Berlin, Berlin, Germany}
%\cortext[cor1]{Corresponding Author}

\journal{Future Generation Computer Systems}

	\begin{frontmatter}
\begin{abstract}
%Due to the volatility of organizational IT landscapes, the number and types of business processes which need to be enacted concurrently may vary to a large extent. 
In business process landscapes, a common challenge is to provide the necessary computational resources to enact the single process steps. %aiming at cost-efficiency while taking into account Quality of Service requirements. 
One well-known approach to solve this issue in a cost-efficient way is to use the notion of elasticity, i.e., to provide cloud-based computational resources in a rapid fashion and to enact the single process steps on these resources. Existing approaches to provide elastic processes are mostly based on Virtual Machines (VMs). % instead of more lightweight software containers. 
Utilizing container technologies could enable a more fine-grained allocation of process steps to computational resources, leading %to a higher level of control over leased computational resources and, thereby, 
to a better resource utilization and improved cost efficiency.

In this paper, we propose an approach to optimize resource allocation for elastic processes by applying a four-fold auto-scaling approach. %, i.e., by combining horizontal and vertical scaling across the VM and container levels. 
The main goal is to minimize the cost of process enactments by using containers. To this end, we formulate and implement a multi-objective optimization problem applying Mixed-Integer Linear Programming and use a transformation step to allocate software services to containers. We thoroughly evaluate the optimization problem and show that it can lead to significant cost savings while maintaining Service Level Agreements, compared to approaches that only use VMs. 
\end{abstract}


\end{frontmatter}

%\flushbottom
%\maketitle
%\thispagestyle{empty}

\section{Introduction}
\label{sec:introduction}
%Necessary only for the first section!
%\IEEEPARstart{C}{loud} computing provides companies with virtually unlimited and rapidly available computational resources in a utility-like fashion~\\citepp{buyya:09}. 
Today\footnote{The accepted manuscript is available at \url{https://doi.org/10.1016/j.future.2022.09.001}.}, cloud-based computational resources are used in many different application areas, e.g., banking~\citep{lampe:13} or healthcare~\citep{ahuja:12}. One particular use case for cloud computing in these areas is the utilization of cloud-based computational resources for the enactment of business processes~\citep{dustdar:11}, i.e., the actual execution and invocation of according software services delivering process functionalities on provisioned computational resources.

Today's business process landscapes are highly volatile and ever-changing, e.g., in terms of the number of arriving process requests or the varying amount of data to be processed~\citep{breu13}. % Also, the boundaries of business process landscapes may frequently change%, e.g., in case of interorganizational business processes are carried out which include new process owners. 
To ensure that these characteristics of business process landscapes are taken into account, Business Process Management Systems~(BPMS) need to make sure that a sufficient amount of computational resources is provided for process enactment, to also fulfill the Quality of Service (QoS) requirements of process owners~\citep{dustdar:11}. 

The utilization of cloud-based computational resources for process enactment leads to the notion of \emph{elastic processes}. For the realization of elastic processes, it is necessary to provide elastic BPMS~(eBPMS), i.e., BPMS which control a process landscape and cloud-based computational resources. eBPMS need to be able to analyze the cost, quality, and resource constraints of a process landscape as well as to monitor the computational resources in use. Based on this information, an eBPMS is capable to scale the computational resources up or down, in or out, based on the current demands of a process landscape~\citep{schulte:15}. 

So far, eBPMS have mostly aimed at the utilization of cloud-based computational resources available as Virtual Machines (VMs), e.g.,~\citep{xu:09, juhnke:11, rosinosky:17}. However, with the advent of container technologies like Docker~\citep{pahl:15}, there is today an alternative approach to utilize cloud-based computational resources for business processes. In fact, containers provide a number of benefits for scheduling and resource allocation, compared to VMs~\citep{xu:14}: First, containers do not virtualize a full operating system and are therefore relatively lightweight. Accordingly, containers can be created significantly quicker than VMs. Second, containers do not need a full, time-consuming reboot, when being reconfigured. Again, this saves time. Third, containers offer flexible configuration possibilities. While cloud providers mostly offer a predefined set of VM instance types, this is not the case for containers. 
This allows a more fine-grained control over computational resources, which may lead to less wasted cost because of unused computational resources. %Fourth, containers allow to implement a DevOps~\citep{bass:15} approach and abstract from inconsistencies of underlying platforms, by unifying the development, test, and production environment. 

When taking into account container-based elastic processes, existing solutions for resource allocation and task scheduling for VM-based elastic processes are not an option any longer: While existing approaches VM-based elastic processes solely rely on horizontal scaling, i.e., leasing and releasing VM instances, e.g.,~\cite{hoenisch:16}, container-based elastic processes also allow to do vertical scaling, i.e., to change the amount of computing power of individual system components~\citep{michael:07}. Hence, task scheduling and resource allocation need to be considered both between the containers and the software services used to enact process steps, and between the containers and VMs. 

Therefore, in this paper, we propose an approach for the cost-efficient auto-scaling of container-based elastic processes which is able to take into account these requirements. Since elastic processes are part of potentially very large process landscapes, optimal resource allocation and process scheduling is a complex task. Finding an optimal solution requires defining and solving a formal optimization model. For this, we conceptualize an optimization model using Mixed-Integer Linear Programming~(MILP). The aim of the model is to find a cost-efficient enactment plan in terms of process step (i.e., task) scheduling and allocation of cloud-based computational resources. A subsequent transformation step is then used to allocate software services to containers. %Our work significantly extends the state of the art (see Section~\ref{sec:related}) by at the same time (i)~Providing an optimal solution, (ii)~Taking into account resource and task sharing between concurrent process instances, and (iii)~considering user-defined Service Level Agreements (SLAs), most importantly process deadlines. 
In brief, the contributions of this paper can be summarized as follows:
\begin{itemize}
	\item We define a system model for elastic processes, including processes, VMs, containers, and services. 
	\item We present the \emph{Four-fold Service Instance Placement Problem} (FFSIPP), an optimization model aiming at minimizing the total cost arising from enacting elastic processes on containers. The optimization model takes 
	into account vertical and horizontal scaling of both VMs and containers in a system landscape. 
	\item We evaluate the prototypical implementation. The evaluation assesses the cost efficiency of the proposed approach as well as the adherence to user-defined Service Level Agreements (SLAs).
\end{itemize}
The remainder of this paper is organized as follows: In Section~\ref{sec:background}, we provide background information about elastic processes and some preliminaries for the subsequent sections. In Section~\ref{sec:systemmodel}, we define the system model necessary to conduct the optimization, while in Section~\ref{sec:problem}, we formulate the multi-objective optimization problem and present a transformation step. We thoroughly evaluate the proposed solution in Section~\ref{sec:evaluation} and discuss related work in Section~\ref{sec:related}. Finally, we conclude this paper in Section~\ref{sec:conclusion}.
\section{Background}
\label{sec:background}
\subsection{Elastic Processes}
\label{sub:elastic}
In brief, an elastic process is a business process that is carried out using elastic cloud infrastructures and resources~\citep{schulte:15}. Hence, elastic processes combine concepts from the field of Business Process Management (BPM) with those of cloud computing, while especially taking advantage of the elasticity features enabled by the resource allocation and virtualization capabilities of the cloud. %For the realization of elastic processes, business processes can be flexibly composed of a number of software services, enacted in scalable cloud infrastructure. 

%In general, elasticity is a concept well-known in areas like physics or economics. %~\citep{tellis:88,biot:55}.
%It describes the sensitivity of one variable in response to the change of some other variable(s)~\citep{herbst:13}. 
With regard to elastic processes, different elasticity properties can be regarded~\citep{dustdar:11}: \emph{resource elasticity}, which describes the change in the amount of used resources (e.g., CPU, RAM, network bandwidth) based on the incoming demand, \emph{cost elasticity}, which refers to the change of cost with the changing amount, duration, and type of provisioned resources, and \emph{quality elasticity}, which describes the change in the delivered QoS in response to fluctuations in the current load of a system. %With regard to elastic processes, the QoS of the overall process and the QoS of the enactment of a complete business process landscape need to be considered. As processes are usually composed of multiple services, the quality of an elastic process depends on the quality of the component services and their interplay. 

Naturally, the three discussed elasticity properties are highly interdependent and changes to one of the properties come with trade-offs for the other properties. For instance, the achieved QoS is often in direct relationship to the utilized resources, which again have an effect on the cost. However, these relationships are not necessarily linear. If these trade-offs are not regarded, the behavior of a business process landscape can lead to unexpected cost or inefficient resource utilization~\citep{kossmann:10}. It is the task of an eBPMS to circumvent such issues and to make sure that elastic processes are enacted in an efficient way while taking into account the QoS demands of the process owners.

\subsection{Preliminaries}
\label{sub:preliminaries}
In the following paragraphs, we discuss the terminology and basic assumptions used within the work at hand. 

In elastic process scenarios, process owners are able to define a \emph{process model}. Process models are composed out of different \emph{process steps}, i.e., single tasks. Each process step can be enacted by a single software \emph{service}. A service can be used in multiple different process models. We explicitly only regard software-based services in our work, i.e., human-provided process steps are not taken into account.

A process model can be requested by a process owner, leading to an executable \emph{process instance}, which features an SLA. Process models can comprise different  workflow patterns~\citep{aalst:03}, e.g., sequences, AND-splits, AND-joins, XOR-splits, XOR-joins, and loops. We assume that the next step of a process instance is known as soon as the preceding step is scheduled. 

In the SLA, non-functional requirements are defined, which need to be taken into account during process enactment, i.e., the invocation of the single software services representing the process steps. Most importantly, an SLA defines the deadline of a process and therefore the priority of the process. If the eBPMS is not able to consider an SLA for a process request, penalty cost accrue, i.e., the process owner gets a penalty fee for the delay.

When a service is deployed on a cloud-based infrastructure, we obtain a \emph{service instance}. Service instances are hosted in the cloud in an isolated way, and more than one instance of the same service may exist at the same time on different cloud resources. A \emph{service invocation} is the unique invocation of a service instance to serve a requested process instance. A service instance can serve multiple incoming service invocations concurrently, as long as it has sufficient computational resources. Thus, multiple process instances of the same or different process models may share service instances for single process steps.

%Process requests may be reoccurring, such that an eBPMS can plan for them in advance, while other process request may follow highly fluctuating and unpredictable request patterns. 
Since the enactment of a varying number of process instances can be requested at any time, a business process landscape can become highly dynamic. An ever-changing process landscape requires different amounts and types of computational resources during different time periods~\citep{dustdar:11}. In the work at hand, we investigate the usage of cloud-based software containers for the provisioning of these computational resources. We focus on the actual process enactment time, i.e., the execution time for the single services, and assume that communication times, e.g., between the public and the private cloud, as well as data transfer times, are negligible. In addition, we assume that the services are primarily computing-intensive. 

In order to take care of the dynamics in a process landscape, the leased cloud-based computational resources need to be optimized for obtaining a cost- and resource-efficient system landscape. For this, we aim at controlling cloud-based computational resources both at the container and VM level in order to avoid a waste of resources and therefore unnecessary cost.

The cloud-based computational resources can be leased from both private and public cloud providers. We assume that both private and public clouds offer different \emph{VM types}. Once leased, the resources of a running \emph{VM instance} of a certain type cannot be extended during a particular Billing Time Unit~(BTU). A BTU defines the minimum leasing duration of a VM instance. An arbitrary number of VM instances with arbitrary resource configurations can be leased at any time. Multiple \emph{containers} can be deployed on different leased VM instances. These containers host the software services necessary to enact a business process in terms of the single process steps. Since we use Docker as container technology, it is possible to create a Docker container hosting a particular service by deploying a Docker image on leased VM instances. The Docker image is downloaded from an according Docker registry. Once an image has been downloaded, it stays available on a VM instance, i.e., does not need to be downloaded again. Docker containers for the same software service can be deployed on arbitrary many VM instances, but each VM instance can only run Docker containers of different service types. Each container can be assigned with a varying amount of computational resources in terms of CPU and RAM. The assigned resources of a container can be adapted on demand and are only limited by the available resources of the underlying VM instance.

In order to provide these functionalities, an eBPMS needs to manage cloud-based computational resources (i.e., lease and release VM instances), to deploy the necessary amount and types of containers with invokable software services on the leased resources, and to schedule incoming process requests (in terms of service invocations) on the available service instances. This has to be done in a dynamic fashion, as new process requests arrive, the process landscape may change, and new process steps have to be scheduled. Since an eBPMS needs to control both the cloud and process landscape, it acts as a middleware, which provides Platform-as-a-Service (PaaS) functionalities to process owners. At the same time, the eBPMS acts as the allocator and controller of cloud-based computational resources. This means that the entity controlling the process execution at the same time needs to be able to control services, and utilize cloud resources. Therefore, we do not take into account services offered in a Software-as-a-Service (SaaS) manner. Instead, the eBPMS controls the deployment and invocation of the service instances. 

%At the Software-as-a-Service (SaaS) level, an eBPMS needs to be able to deploy the single software services of elastic processes, and to make sure that the services have sufficient computational resources. 

In this paper, we present an approach to compute an optimal resource allocation and task scheduling plan, by providing a solution based on MILP. In theory, each eBPMS that is able to provide the functionalities discussed in the last paragraph can be used together with the proposed solution. Within the presented work, we make use of an extended version of the Vienna Platform for Elastic Processes (ViePEP), which is a fully-fledged research eBPMS~\citep{hoenisch:16,WHS+19}. %ViePEP provides the possibility to provide self-adaptive process enactment by following the MAKE-K cycle~\citep{kephart:03}.

\section{Basic Approach and System Model}
\label{sec:systemmodel}
\subsection{Basic Approach}
\label{sub:basicapproach}
As outlined in Section~\ref{sec:introduction}, we provide a four-fold auto-scaling approach for elastic processes. This means that both VMs and containers are adjusted horizontally and vertically.

Horizontal scaling of VMs is accomplished by changing the number of leased VM instances of the system, while horizontal scaling of containers is realized by changing the number of deployed container instances offering a certain software service.

Vertical scaling of VMs refers to the change of computational resources of a system, while the number of deployed VM instances remains unchanged. It has to be noted that vertical scaling in this context does not refer to single VM instances, but to the system landscape as a whole, as we do not address problems related to reassigning computational resources to running VM instances, but allow to exchange running VM instances by VM instances of different types offering different resources.

Vertical scaling of containers is realized by resizing deployed container instances so that they offer more or less computational resources of the underlying VM instance. This resizing is achieved by changing the assigned CPU and memory shares to container instances.

%Our approach is generally focusing on managing computational resources in terms of CPU and memory resources, but can be adjusted to take additional resource constraints, e.g., GPU resources, into consideration.

\subsection{System Model}
\label{sub:systemmodel}
In order to define our system model, we first describe the variables used to define the processes in Section~\ref{subsub:process}. Afterwards, we present the necessary time variables in Section~\ref{subsub:time} and formally define VMs (Section~\ref{subsub:vms}), as well as containers and services (Section~\ref{subsub:containers}). Finally, we introduce further variables of the system model in Section~\ref{subsub:further}. An overview of all applied variables is given in Tables~\ref{tab:notation}-\ref{tab:notation3}. 

\subsubsection{Process Variables}
\label{subsub:process}
\begin{table}[h]
	\caption{Process Variables}
	\label{tab:notation}
	\small
\begin{tabularx}{\columnwidth}{r|X}
$P = \left\{1, \ldots, p^{\#}\right\}$& Set of process models $p \in P$\\
$I_{p} = \left\{1, \ldots, i_{p}^{\#}\right\}$ & Set of process instances $i_{p} \in I_{p}$, with $p \in P$ indicating which model is used for a particular process instance\\
$J_{i_{p}} = \left\{1, \ldots, j_{i_{p}}^{\#}\right\}$ & Set of process steps $j_{i_{p}} \in i_{p}$ not yet enacted, with $i_{p}$ indicating the process instance of the step\\
$J_{i_{p}}^{*} \subseteq J_{i_{p}}$ & Set of \emph{next} process steps $j_{i_{p}}^{*} \in J_{i_{p}}^{*}$ to be  enacted\\
$J_{i_{p}}^{run} \not\subset J_{i_{p}}$ & Set of currently enacted (i.e., running) process steps $j_{i_{p}}^{run} \in J_{i_{p}}^{run}$\\
$DL_{i_{p}}$ & Deadline for the enactment of $i_{p}$\\
$DL_{j_{i_{p}}^{*}}$ & Deadline for starting the enactment of a next process step $j_{i_{p}}^{*}$\\
$r_{j_{i_{p}}}^{C}, r_{j_{i_{p}}}^{R}$ & Resource demands of a process step $j \in i_{p}$
in terms of CPU ($r_{j_{i_{p}}}^{C}$) and RAM ($r_{j_{i_{p}}}^{R}$)\\
\end{tabularx}
\end{table}

Business process landscapes are made up from a potentially very large amount of process instances, which stem from different process models. Therefore, we consider multiple process models, with $P = \left\{1, \ldots, p^{\#}\right\}$ representing the set of process models, and $p \in P$ indicating a specific process model. In order to differentiate between different process models, $p$ (as well as the other variables used in order to describe processes, VMs, containers, etc.) take sequential integer values. The set of process instances of a specific process model $p$ is represented by $I_{p}$, while $i_{p} \in I_{p}$ indicates a specific process instance of the process model~$p$. 

Process models (and accordingly process instances) are composed of different process steps. The set of process steps not yet enacted (and not yet started) to realize a specific process instance $i_{p}$ is represented by $J_{i_{p}}$, while $j_{i_{p}} \in J_{i_{p}}$ refers to a specific process step of a process instance $i_{p}$. The set of \textit{next} process steps (i.e., the immediately next steps in a particular process instance) that have to be scheduled for enactment to realize a specific process instance $i_{p}$ is represented by $J_{i_{p}}^{*}$, while $j_{i_{p}}^{*} \in J_{i_{p}}^{*} \subseteq J_{i_{p}}$ refers to a specific next process step of the process instance~$i_{p}$. %	Note: AND-Constructs can execute multiple paths in parallel and can, therefore, have multiple next process steps in the set $J_{i_{p}}^{*}$.
$j_{i_{p}}^{run} \in J_{i_{p}}^{run} \not\subset J_{i_{p}}$ is a specific currently running process step of the process instance $i_{p}$, with the set of currently running and not yet finished process steps of a specific process instance $i_{p}$ being represented by  $J_{i_{p}}^{run}$.  %Note: AND-Constructs can execute multiple paths in parallel and can, therefore, have multiple currently running process steps in the set $J_{i_{p}}^{run}$.\\

As written above, process owners are able to define a deadline until when a particular requested process instance needs to be finished. The deadline for the enactment of the process instance $i_{p}$ is defined by $DL_{i_{p}}$ and refers to a certain point in time. The deadline for starting the enactment of a next process step $j_{i_{p}}^{*}$ such that the process instance still meets its process deadline $DL_{i_{p}}$, while performing a worst-case analysis considering all subsequent steps, is defined by $DL_{j_{i_{p}}^{*}}$, and refers also to a certain point in time. The overall starting time of a process instance is defined by its first process step. We discuss the according computations in Section~\ref{sub:constraints}.

Finally, the resource demands of a process step $j$ of the process instance $i_{p}$ in terms of CPU and RAM are indicated by the respective terms $r_{j_{i_{p}}}^{C}$ (CPU) and $r_{j_{i_{p}}}^{R}$ (RAM).

% Note: Den Gewichtungsfaktor führe ich später ein, wenn dieser erklärt wird.
%$\omega_{DL}$
%	& The weighting factor for controlling the effect of deadline constraints is
%	indicated by $\omega_{DL}$.\\

\subsubsection{Time Variables}
\label{subsub:time}
\begin{table}[h]
	\small
	\caption{Time-related Variables}
	\label{tab:notationTime}
	\begin{tabularx}{\columnwidth}{r|X}
		$t$ & A specific point in time\\
		$\tau_{t}$ & Current time period starting at $t$\\
		$\tau_{t+1} $ & Next time period, starting at $t+1$\\
		$ t_{j_{i_{p}}}$ & Point in time $j_{i_{p}}$ is scheduled\\
		$\epsilon$ & Minimum time period between two
		optimization runs\\
	\end{tabularx}
\end{table}	
The planned optimization of resource allocation and task (i.e., process step) scheduling is done for a particular time period $\tau_{t}$, i.e., the optimization is carried out for this time period and subsequently repeated for each new time period. $\tau_{t+1}$ refers to the next time period starting at $t+1$. At the beginning of each time period, the optimization model is calculated based on all currently available information and its results are enacted. How often the optimization is carried out depends on the volatility of a business process landscape as well as the demands of the BPMS operator (see Section~\ref{sec:problem}). %In the most extreme case, a new optimization could, e.g., be carried out whenever a new process request arrives or a service is delayed. In other cases, the optimization might be done in predefined intervals.
%Accordingly, the parameter $t$ is used to describe the point in time that marks the beginning of a time period. 

The point in time at which a specific service invocation $j_{i_{p}}$ is scheduled is referred to as $t_{j_{i_{p}}}$.  Finally, $\epsilon$ describes (in milliseconds) the minimum time period that is at least needed between two optimization runs, e.g., in order to cater for the amount of time the optimization itself needs, or in order to avoid too frequent changes. This parameter can be set manually by a system operator.

\subsubsection{Virtual Machine Variables}
\label{subsub:vms}
\begin{table}[h]
	\caption{VM Variables}
	\label{tab:notationVMs}
	\small
	\begin{tabularx}{\columnwidth}{r|X}
		$V = \left\{1, \ldots, v^{\#}\right\}$ & Set of VM types $v \in V$\\
		$K_{v} = \left\{1, \ldots, k_{v}^{\#}\right\}$ & Set of leasable VM instances $k_{v} \in K_{v}$ of type~$v$\\
		$BTU_{v}$ & BTU of $k_{v}$\\
		$BTU^{max}$ & Arbitrary large upper bound of possible leasable BTUs for any
		$k_{v}$\\
		$c_{v}$ & Leasing cost for VM instances of type $v$ for one full BTU \\
		$\beta_{(k_{v}, t)} \in \left\{0,1\right\}$
		& Indicates if $k_{v}$ is/was already running at the beginning of $\tau_{t}$ \\
		$g_{(k_{v}, t)} \in \left\{0,1\right\}$
		& Indicates if $k_{v}$ was already running at the beginning of $\tau_{t}$ or is leased
		during $\tau_{t}$\\
		$d_{(k_{v}, t)}$ & Remaining leasing duration of $k_{v}$ at the beginning of $\tau_{t}$\\
		$s_{v}^{C}, s_{v}^{R}$ & Resource supplies of a VM type $v$ in terms of CPU ($s_{v}^{C}$)
		and RAM ($s_{v}^{R}$)\\
		$f_{k_{v}}^{C}, f_{k_{v}}^{R}$
		& Free resources in terms of CPU ($f_{k_{v}}^{C}$) and RAM
		($f_{k_{v}}^{R}$) of $k_{v}$\\
	\end{tabularx}
\end{table}

The presented approach allows the utilization of VMs from multiple public and private cloud providers, which both may provide VMs in different configurations and at different cost. The set of VM types is represented by $V$, while $v~\in~V$ indicates a specific VM type. We assume that the number of leasable VM instances in the public cloud is virtually unlimited~\citep{mell:11}, but a private cloud provider may only possess limited computational resources which can be leased. %In our system model, computational resources can be leased from both public and private clouds. 
The set of leasable VM instances of type $v$ is represented by $K_{v}$, while $k_{v} \in K_{v}$ is the $k^{th}$ VM instance of type $v$. %Note: For a time period starting at $t$, we assume the number of leasable VMs of type $v$ to be limited with $k_{v}^{\#}$. 

The BTU of a VM instance $k_{v}$ of VM type $v$ is indicated by $BTU_{v}$. A BTU is a billing cycle and the minimum leasing duration for a VM instance $k_{v}$. $c_{v}$ represents the leasing cost for VM instances of type $v$ for one full BTU. Especially if the BTU defined by a cloud provider is relatively short, one BTU will not be enough to cover a complete optimization time period $\tau_{t}$. Therefore, we define the decision variable $\gamma_{(v,t)}~\in~\mathbb{N}_{0}^{+}$, which indicates the total number of BTUs to lease any VMs of type $v$ in the time period $\tau_{t}$. %Releasing a VM before the end of a BTU leads to a waste of already paid resources.
 An arbitrary large upper bound of possible leasable BTUs for any VM instance $k_{v}$ is expressed by the helper variable $BTU^{max}$. This upper bound is needed in order for the computation of the optimization problem to not getting stuck in a loop.%, and is therefore predefined.

The helper variable $\beta_{(k_{v}, t)} \in \left\{0,1\right\}$ contributes to calculating $g_{(k_{v}, t)} \in \left\{0,1\right\}$. $g_{(k_{v}, t)}$ indicates if the VM instance $k_{v}$ has already been running at the beginning of time period~$\tau_{t}$ or is leased during $\tau_{t}$. $g_{(k_{v}, t)}$ therefore helps identifying the VM instances $k_{v}$ that are either already running or are being newly leased in $\tau_{t}$. This is important in order to reuse already running $k_{v}$ instead of spawning new ones. 
$d_{(k_{v}, t)}$ indicates the remaining leasing duration of the VM instance $k_{v}$ at the beginning of the time period $\tau_{t}$. These variables are important in order for the optimization model to be able to calculate how many computational resources are available for how long.

%The weighting factor for controlling the effect of the remaining leasing duration on the optimization outcome is indicated by $\omega_{d}$.
Finally, the resource supplies of a VM type $v$ in terms of CPU and RAM are indicated by $s_{v}^{C}$ (CPU) and $s_{v}^{R}$ (RAM). In addition, the unused resources of a VM instance $k_{v}$ are indicated by $f_{k_{v}}^{C}$ (CPU) and $f_{k_{v}}^{R}$~(RAM). 

\subsubsection{Container and Service Variables}
\label{subsub:containers}
\begin{table}[h]
	\caption{Container Variables}
	\label{tab:notationContainers}
	\small
	\begin{tabularx}{\columnwidth}{r|X}
		$ST = \left\{1, \ldots, st^{\#}\right\}$ & The set of all container types mapping all possible service types $st$\\
		$st_{j}$ & Service type of a certain process step $j$\\
		$C_{st} = \left\{1, \ldots, c_{st}^{\#}\right\}$ & Set of container instances $c_{st} \in C_{st}$ mapping specific service instances of type $st$ \\		
%		$c_{{st}_j}$ & Container instance of type $st$ that can run $j$\\		
		$z_{(st, k_{v}, t)} \in \left\{0,1\right\}$ & Indicates if any containers of type $st$ have already
		been deployed on $k_{v}$ until $t$\\
		$M$ & Arbitrary large upper bound of possible container deployments for any $k_{v}$\\
		$N$ & Arbitrary large upper bound of possible service invocations for any $c_{st}$\\
	\end{tabularx}
\end{table}

Following the four-fold auto-scaling approach discussed in Section~\ref{sub:basicapproach}, containers are deployed on leased VMs. Therefore, in addition to the VM variables described in the last subsection, we also need to formally define the containers which are used to host and invoke the actual software services. 

The set of all container types mapping all possible service types is represented by $ST$, while $st~\in~ST$ indicates a specific container type for a certain service type. Notably, the container type defines which service types (i.e., images) can be offered by a container instance.
The service type of a certain process step $j$ is indicated by $st_{j}$, i.e., a process step can only be executed by the according service type.

The set of container instances mapping specific service instances of type $st$ is represented by $C_{st}$, while $c_{st}~\in~C_{st}$ is the $c^{th}$ instance of type $st$. A container instance of a certain service type $st$ that can run a process step $j$ is given by $c_{st_{j}}$. 

The variable $z_{(st, k_{v}, t)} \in \left\{0,1\right\}$ indicates if any containers of service type $st$ %, i.e., any $c_{st} \in C_{st}$
have already been deployed on a VM instance $k_{v}$ (and hence the container image for starting new container instances of service type $st$ has been downloaded on $k_{v}$) until $t$. This information is needed, since downloading an image takes time and should therefore be avoided (see Section~\ref{subsub:further}). 

%The weighting factor for controlling the effect of existing images on VM instances for the optimization outcome is indicated by $\omega_{z}$.

Finally, two variables are introduced in order to make sure that the optimization is not getting stuck in a loop: 
An arbitrary large upper bound of possible container deployments for any VM instance $k_{v}$ is expressed by the helper variable $M$, and an arbitrary large upper bound of possible service invocations for any container instance $c_{st}$ is expressed by the helper variable $N$. Again, these parameters are predefined.

%	$\omega_{s}^{C}, \omega_{s}^{R}$
%& Weighting factors for controlling the effect of resource supply of container instances in terms of CPU ($\omega_{s}^{C}$) and RAM  ($\omega_{s}^{R}$) on the optimization outcome are indicated by the respective terms.\\

\subsubsection{Further Variables}
\label{subsub:further}
\begin{table}[h]
	\small
	\caption{Further Variables}
	\label{tab:notationFurther}
	\begin{tabularx}{\columnwidth}{r|X}
		$\Delta_{v}$
		& Time to start a new VM of type $v$\\
		$\Delta$ = $max_{v \in
			V}(\Delta_{v})$ & Maximum VM-startup time of any type $v$\\
		$\Delta_{st}$ & Time for pulling a container image for $st$ on a
		VM instance for the first time\\
		$\Delta_{c_{st}}$ & Time for starting a container instance from an existing image\\
		$e_{i_{p}}$ & Remaining enactment duration of $i_{p}$\\
		$e_{i_{p}}^{seq}, e_{i_{p}}^{L_{a}}, e_{i_{p}}^{L_{x}}, e_{i_{p}}^{RL}$ &
		Remaining enactment duration of sequences ($e_{i_{p}}^{seq}$),
		AND-blocks ($e_{i_{p}}^{L_{a}}$), XOR-blocks ($e_{i_{p}}^{L_{x}}$) and repeat
		loops ($e_{i_{p}}^{RL}$) of $i_{p}$\\
		$L = \left\{1 \ldots l^{\#}\right\}$ & Set of all possible paths $l$ in $i_p$ \\
		$L_{a} \cup L_{x} \cup RL \subseteq L$ & Set of possible paths for AND-blocks $L_{a}$, XOR-blocks $L_{x}$, and repeat loops $RL$\\
		$re$ & Maximum repetitions of a repeat loop\\
		$\hat{e}_{i_{p}}^{s}, \hat{e}_{i_{p}}^{l}$ 	& Remaining combined service enactment duration, time for pulling needed container images, container deployment and VM startup time for all
		not yet scheduled service invocations of sequences or repeat loop blocks
		($\hat{e}_{i_{p}}^{s}$) and paths in AND- or XOR-blocks
		($\hat{e}_{i_{p}}^{l}$) that still need to be invoked to finalize $i_{p}$\\ 
		$ex_{j_{i_{p}}^{run}}$ 	& Remaining enactment time of process steps already scheduled in previous
		periods and not finished until the start of $\tau_{t}$\\
		$ex_{j_{i_{p}}^{*}}$ & Combined enactment duration, container deployment and VM-startup
time of $j_{i_{p}}^{*}$\\
		$e_{i_{p}}^{p}$ & Penalty time units of $i_{p}$, i.e., the enactment that occurs beyond the deadline $DL_{i_{p}}$\\
		$c_{i_{p}}^{p}$ & The penalty cost per time unit of delay of $i_{p}$\\
	\end{tabularx}
\end{table}

Finally, some variables need to be introduced as part of the system model in order to cater for start, deployment, enactment, and penalty times of VMs, containers, and services. 

To start with, the time in milliseconds to start a new VM of type $v$ is indicated by $\Delta_{v}$. The maximum VM-startup time of any type, i.e., $max_{v \in V}(\Delta_{v})$ is expressed by $\Delta$. This variable is later on needed for the worst-case estimation we perform during optimization (see Section~\ref{sub:constraints}). The time for pulling a container image for a service type $st$ on a
VM instance for the first time is expressed by $\Delta_{st}$. The time for starting a container instance from an existing image for a
service type $st$ on a VM instance is expressed by $\Delta_{c_{st}}$

The remaining enactment time of a process instance~$i_{p}$ is expressed by $e_{i_{p}}$. $e_{i_{p}}$ does not include the enactment and deployment time of the process steps that will be scheduled for the current optimization period $\tau_{t}$, nor the remaining enactment time of the currently already running steps. 

To calculate $e_{i_{p}}$, we need to separately calculate the remaining enactment time of sequences, AND-blocks, XOR-blocks, and repeat loops in a process model. These are given by $e_{i_{p}}^{seq}$ (sequences), $e_{i_{p}}^{L_{a}}$~(AND-blocks), $e_{i_{p}}^{L_{x}}$ (XOR-blocks), and $e_{i_{p}}^{RL}$ (repeat loops). These terms do not include the remaining enactment time of the currently running steps nor the enactment and deployment times for process steps that will be scheduled for the current optimization period~$\tau_{t}$. 

To calculate the remaining enactment times for the mentioned workflow patterns, it is necessary to know the potential paths (i.e., process steps) in a process instance. The set of all paths is represented by $L = \left\{1 \ldots l^{\#}\right\}$, while $l \in L$ indicates a specific path within a process. The set of possible paths for AND-blocks $L_{a}$, XOR-blocks $L_{x}$ and repeat loop constructs $RL$ are expressed by the respective terms, with $L_{a} \cup L_{x} \cup RL \subseteq L$. For the repeat loops, again, we have to introduce a helper variable, which provides an upper limit. Accordingly, the maximum repetitions of a repeat loop are indicated by $re$. %Notably, $re$ can be arbitrarily large.

As written above, the remaining enactment duration of a process instance ($e_{i_{p}}$) is limited to the actual service enactment times. The remaining combined service enactment duration, time for pulling needed container images, container deployment and VM startup time for all not yet scheduled service invocations of sequences or repeat loop blocks ($\hat{e}_{i_{p}}^{s}$) and paths in AND- or XOR-blocks ($\hat{e}_{i_{p}}^{l}$) that still need to be invoked to finalize the enactment of the process instance $i_{p}$ are expressed by the respective terms. Still, those values do not include the remaining enactment time of already running process steps ($j_{i_{p}}^{run}$) or process steps that are being scheduled in the current optimization period~$\tau_{t}$. These are provided separately by $ex_{j_{i_{p}}^{run}}$.

The combined enactment duration, container deployment and VM-startup time of the next to be scheduled process step $j_{i_{p}}^{*}$ is expressed by $ex_{j_{i_{p}}^{*}}$.

To take into account penalty cost which may accrue if a process instance does not meet its specified deadline, two further parameters are introduced. First, the  penalty time of a process instance $i_{p}$ measured in time units, i.e., the enactment duration of ${i_{p}}$ that occurs beyond the deadline $DL_{i_{p}}$ of $i_{p}$, is expressed by $e_{i_{p}}^{p}$. This variable is used by the optimization model in order to calculate if it is cheaper to pay the penalty fee than to lease further computational resources. We assume linear penalty fees, i.e., the fees grow with the delay of process enactment. The penalty cost per time unit of delay of the process instance $i_{p}$ are represented by $c_{i_{p}}^{p}$.%\vfill\null
%\newpage

We allow a system operator to weight the different parts of the objective function presented in Section~\ref{sec:problem}, e.g., to give a higher importance to the utilization of leased computational resources. This allows to control the effects of different constraints on the optimization outcome. These weighting factors are shown in Table~\ref{tab:notationWeight}. Their application is explained in more detail in Section~\ref{sub:objective}.
\begin{table}[t]
	\caption{Weighting Factors}
	\label{tab:notationWeight}
	\small
	\begin{tabularx}{\columnwidth}{r|X}
		$\omega_{DL}$ & Weighting factor for controlling the effect of deadline constraints\\
		$\omega_{d}$
		& Weighting factor for controlling the effect of the remaining leasing 
		duration on the optimization outcome\\
		$\omega_{f}^{C}, \omega_{f}^{R}$
		& Weighting factors for controlling the effect of free resources in terms of
		CPU ($\omega_{f}^{C}$) and RAM ($\omega_{f}^{R}$)\\
		$\omega_{z}$ & Weighting factor for controlling the effect of existing images on VM instances \\	
	\end{tabularx}
\end{table}

Finally, Table~\ref{tab:notation3} provides an overview of the decision variables needed in our decision model. $\gamma$ primarily acts as a helper variable and has already been mentioned in Section~\ref{subsub:vms}. $a$, $x$ and $y$ are the core decision variables of our objective function and accordingly further discussed in the next section.
\begin{table}[t]
	\small
	\caption{Decision Variables}
	\label{tab:notation3}
	\begin{tabularx}{\columnwidth}{r|X}
		%		\hline
		%		\multicolumn{2}{c}{Decision Variables}\\
		$\gamma_{(v,t)} \in \mathbb{N}_{0}^{+}$ & Number of BTUs to lease any
		VM of type $v$ in $\tau_{t}$\\
		$a_{(c_{st}, k_{v}, t)} \in \left\{0,1\right\}$&Deployment of $c_{st}$ on $k_v$ in $\tau_{t}$\\
		$x_{(j_{i_{p}}, k_{v}, t)} \in \left\{0,1\right\}$ & Invocation of $j \in i_{p}$ on $k_{v}$ in $\tau_{t}$\\
		$y_{(k_{v}, t)} \in \left\{\mathbb{N}_{0}^{+}\right\}$ &  Number of BTUs to lease $k_{v}$ in $\tau_{t}$\\
	\end{tabularx}
\end{table}

\section{Multi-Objective Problem Formulation}
\label{sec:problem}
Based on the system model introduced in the last section, we are now able to formulate our task scheduling and resource allocation problem FFSIPP as a MILP optimization problem. The multi-objective optimization problem represents the scheduling problem for one optimization round starting at time~$t$. An optimization round is triggered whenever new process steps should be scheduled, for example, whenever new process requests arrive or when previously running steps of process instances are finished. Furthermore, the optimization is triggered when SLA violations can be foreseen. 

Our optimization problem takes complex workflow patterns, penalty cost, container-based metrics, different pricing models, and BTUs for VMs into consideration. The optimization problem focuses on minimizing the cost that arise from enacting elastic process landscapes on container-based cloud infrastructures while taking into account process SLAs, i.e., QoS in terms of deadlines. Solutions for the problem describe when and on which containers and service instances to schedule service invocations, on which VMs to place containers and service instances, and which VM to lease or release.

\subsection{Objective Function}
\label{sub:objective}
{\footnotesize
	\begin{eqnarray*}
		min\Big[ && \sum_{\substack{v \in V}} \big(c_{v} \times \gamma_{(v,t)}\big) \\
		%min &\sum_{\substack{v \in V}} \big(c_{v} \times \gamma_{(v,t)}\big) \\
		%TERM2
		& + &\sum_{\substack{p \in P}} \sum_{\substack{i_{p} \in I_{p}}}
		\big(c_{i_{p}}^{p}  \times e_{i_{p}}^{p}\big) \\
		%TERM3
		& + &\sum_{\substack{p \in P}} \sum_{\substack{i_{p} \in I_{p}}}
		\sum_{\substack{j_{i_{p}}^{*} \in J_{i_{p}}}}
		\sum_{\substack{v \in V}} \sum_{\substack{k_{v} \in K_{v}}}
		\big(\omega_{z} \times (1-z_{(st_{j},k_{v},t)}) \times ~ x_{(j_{i_{p}},k_{v},t)}\big) \\
		%TERM4
		& + &\sum_{\substack{p \in P}} \sum_{\substack{i_{p} \in I_{p}}}
		\sum_{\substack{j_{i_{p}}^{*} \in J_{i_{p}}}}
		\sum_{\substack{v \in V}} \sum_{\substack{k_{v} \in K_{v}}}
		\big(\omega_{d} \times d_{(k_{v}, t)} \times x_{(j_{i_{p}},k_{v},t)}\big)
		\\
		%TERM5
		& + &\sum_{\substack{v \in V}} \sum_{\substack{k_{v} \in K_{v}}}
		\big(\omega_{f}^{C} \times f_{k_{v}}^{C} + \omega_{f}^{R} \times f_{k_{v}}^{R}\big) \\
		%TERM6
		& + &\sum_{\substack{p \in P}} \sum_{\substack{i_{p} \in I_{p}}}
		\sum_{\substack{j_{i_{p}}^{*} \in J_{i_{p}}}}
		\sum_{\substack{v \in V}} \sum_{\substack{k_{v} \in K_{v}}}
		\big(\omega_{DL} \times (DL_{j_{i_{p}}^{*}}-\tau_{t}) \times ~ x_{(j_{i_{p}},k_{v},t)}\big) \Big]
	\end{eqnarray*}
}
The equation above presents the objective function of our multi-objective optimization problem FFSIPP. As we are aiming at cost efficiency, the objective function is subject to minimization under the constraints introduced in Section~\ref{sub:constraints}.

%\begin{table}[h]
%\begin{equation}
%\begin{aligned}
%%TERM1
%&min \sum_{\substack{v \in V}} \big(c_{v} * \gamma_{(v,t)}\big) \\
%%min &\sum_{\substack{v \in V}} \big(c_{v} * \gamma_{(v,t)}\big) \\
%%TERM2
% + &\sum_{\substack{p \in P}} \sum_{\substack{i_{p} \in I_{p}}}
%\big(c_{i_{p}}^{p}*e_{i_{p}}^{p}\big) \\
%%TERM3
% + &\sum_{\substack{p \in P}} \sum_{\substack{i_{p} \in I_{p}}}
%\sum_{\substack{j_{i_{p}}^{*} \in J_{i_{p}}}}
%\sum_{\substack{v \in V}} \sum_{\substack{k_{v} \in K_{v}}}
%\big(\omega_{z} * (1-z_{(st_{j},k_{v},t)}) * x_{(j_{i_{p}},k_{v},t)}\big)
%\\
%%TERM4
% + &\sum_{\substack{p \in P}} \sum_{\substack{i_{p} \in I_{p}}}
%\sum_{\substack{j_{i_{p}}^{*} \in J_{i_{p}}}}
%\sum_{\substack{v \in V}} \sum_{\substack{k_{v} \in K_{v}}}
%\big(\omega_{d} * d_{(k_{v}, t)} * x_{(j_{i_{p}},k_{v},t)}\big)
%\\
%%TERM5
%+ &\sum_{\substack{v \in V}} \sum_{\substack{k_{v} \in K_{v}}}
%\big(\omega_{f}^{C}*f_{k_{v}}^{C} + \omega_{f}^{R}*f_{k_{v}}^{R}\big) \\
%%TERM6
% + &\sum_{\substack{p \in P}} \sum_{\substack{i_{p} \in I_{p}}}
%\sum_{\substack{j_{i_{p}}^{*} \in J_{i_{p}}}}
%\sum_{\substack{v \in V}} \sum_{\substack{k_{v} \in K_{v}}}
%\big(\omega_{DL} * (DL_{j_{i_{p}}^{*}}-\tau_{t})
%* x_{(j_{i_{p}},k_{v},t)}\big)
%\end{aligned}
%\label{eq:obj_new}
%\end{equation}
%\end{table}

The objective function considers six terms for calculating a scheduling plan, indicated by the decision variable $x_{(j_{i_{p}},k_{v},t)}$, which decides on which VM instance $k_{v}$ to schedule a service invocation for process step $j$ of process instance~$i_{p}$ in the time period $\tau_{t}$. In the following paragraphs, the single terms of the objective function are explained in more detail: 
\begin{itemize}
	\item \textbf{Term 1} - \emph{minimize total VM leasing cost:} The 	first term calculates the total cost for leasing VM instances of each VM type $v$ at time $t$, by considering the leasing cost $c_{v}$ for the VM types and the total number of leased BTUs $\gamma_{(v,t)}$ that VM instances of the corresponding type $v$ are leased at $t$. The aim is to minimize the total cost for leasing VMs.
	\item \textbf{Term 2} - \emph{minimize penalty cost:} The second term 
	calculates the total penalty cost that arise from scheduling decisions that
	lead to enactment times for process instances beyond their defined deadlines. The penalty time $e_{i_{p}}^{p}$ of a
	process instance $i_{p}$ is calculated according to a worst-case analysis.
	The aim is to minimize such penalty cost and therefore make sure that
	incoming process requests are enacted in time, as defined in their SLAs. 
	However, it might in some cases be more cost-efficient to pay the penalty fee than to lease additional computational resources.
	\item \textbf{Term 3} - \emph{minimize container deployment time:} The
	third term minimizes the sum of time needed for deploying new container instances on VM instances $k_{v}$.
	Notably, we do not consider directly the scheduling of container instances in the optimization problem (see Section~\ref{sub:transformation}). Instead, we consider the
	placement of service steps on the VM instances and use the service types of
	the scheduled process steps to indirectly calculate the potential deployment time for
	container instances.
	The subterm $(1-z_{(st_{j},k_{v},t)})$ makes sure that the weighting factor $\omega_{z}$ is
	only considered when the cache for container deployment does not exist on the
	VM instance $k_{v}$. As the full term is minimized, the system prefers choosing VM instances
	for assigning service invocations where the cache for deploying the needed container instance already exists. This term accordingly describes a trade-off between the time for deployment and cost. 
	\item \textbf{Term 4} - \emph{maximize future resource supply of already
		leased VMs:} The fourth term aims at maximizing the future resource supply of already leased VM instances, by
	giving a preference to directly scheduling service invocations on VMs with a shorter remaining leasing duration
	$d_{(k_{v}, t)}$ compared to other VMs with already existing longer leasing durations.
	Although at time~$t$ this preference does not lead to any cost-based advantages, it assures that the already leased
	resources offer more long-running resources for future optimization periods, which can then be used by services. This reduces the amount of wasted computational resources. 
	The term calculates the remaining leasing duration of all VMs using the 
	weighting factor $\omega_{d}$. As we want to give a preference to deploying containers on
	VM instances with a small leasing duration, we minimize the term to
	maximize the future remaining leasing durations of available resources.
	\item \textbf{Term 5} - \emph{minimize sum of unused present VM
		resources:} The fifth term minimizes the sum of all leased but unused
	computational resource capacities in terms of CPU ($f_{k_{v}}^{C}$) and RAM
	($f_{k_{v}}^{R}$) of all leased VM instances. Due to this term, our
	approach makes sure to not only schedule all service invocations that cannot
	be delayed any further without leading to penalty cost, but to also
	pre-schedule service invocations that may already be enacted but have a
	distant deadline. Again, this leads to a minimization of wasted computational resources, and therefore contributes indirectly to the cost optimization. We consider this term using weighting factors for CPU~($\omega_{f}^{C}$) 	
	and RAM ($\omega_{f}^{R}$)
	resources.%, such that we are able to give more importance to the first two terms of our objective function.
	\item \textbf{Term 6} - \emph{maximize the importance of scheduled service
		invocations:} The last term calculates the difference between the last possible
	scheduling deadline of the next schedulable steps when performing a worst-case
	analysis based on the current time. The term considers all process instances
	and aims at giving a scheduling preference to process steps with closer deadlines. The rationale behind this is that process steps with such deadlines could lead to an SLA breach and according penalty cost. The term is responsible for defining the importance of schedulable process
	steps that do not necessarily have a pressing process deadline on remaining
	free resources.
	We consider
	the actual enactment deadline for the next schedulable steps based on the
	worst-case analysis instead of simply comparing the overall process deadlines.
	We also define the term such that past process deadlines can be considered and
	steps that are past their deadline receive a higher scheduling importance.
	We consider a weighting factor $\omega_{DL}$ to assign a weight to the
	term.
\end{itemize}
In general, the mentioned weighting factors for the single terms of the objective function allow the user (e.g., an eBPMS operator) to control the effect of the respective terms on the overall optimization outcome. For instance, in some settings, $\omega_{DL}$ might be weighted very high in order to support that all process instances are carried out in time. In other settings, this might not be very important, but a system operator might increase $\omega_{f}^{C}$ to minimize unused CPU resources (Term~5) in order to save energy. 

\subsection{Constraints}
\label{sub:constraints}
Apart from the objective function, it is also necessary to define constraints for the FFSIPP. %These constraints are taken into account when leasing computational resources and doing the scheduling for elastic processes. 
To start with, Constraint (\ref{eq:constr2_new}) makes sure that deadlines are considered for each
process instance $i_{p}$.
The constraint does not guarantee that deadlines are not violated, but it defines
the penalty times $e_{i_{p}}^{p}$ of process instances which are subject to
minimization in the presented \emph{Term~2} of the objective function.
The constraint demands that the current time $\tau_{t}$ plus the remaining worst-case
enactment time of process instances $ex_{j_{i_{p}}}^{run} + ex_{j_{i_{p}}}^{*} + e_{i_{p}}$
is smaller or equal to the defined deadline $DL_{i_{p}}$ 
plus the potential penalty time~$e_{i_{p}}^{p}$ which occurs when process instances finish after
their deadline. This constraint considers the remaining enactment time including
the time of process steps that are being scheduled during the current time
period and process steps that have already been scheduled in previous periods
and are still running. Naturally, the starting time for a complete process instance is defined by the computed starting time of the very first process step within the process model. For this, different workflow patterns are taken into account (see below).
\begin{eqnarray}
\tau_{t} + ex_{j_{i_{p}}}^{run} + ex_{j_{i_{p}}}^{*} + e_{i_{p}}
\leqslant DL_{i_{p}} + e_{i_{p}}^{p}
\label{eq:constr2_new}
\end{eqnarray}
Constraint (\ref{eq:constr3_new}) helps to define the start of the next
optimization period $\tau_{t+1}$, by demanding that the start of the next
optimization period $\tau_{t+1}$ plus the remaining worst-case enactment time of process instances $e_{i_{p}}$
is smaller or equal to the respectively defined process deadlines $DL_{i_{p}}$ plus the
possible penalty times $e_{i_{p}}^{p}$. %It should be noted that in contrast to related work \citep{hoenisch:16} our approach only considers the worst-case execution time of all process steps that have not been scheduled yet until $\tau_{t}$ and have to be scheduled in a later optimization period $\tau_{t+1}$ or later for the calculation of the start time of the next optimization period $\tau_{t+1}$.
At this stage, the approach explicitly does not consider the enactment time of
currently running process steps or process steps that are scheduled for enactment in
$\tau_{t}$, as the earliest possible time when the next steps of a process instance
can be scheduled is when the previous steps, already running or scheduled in
$\tau_{t}$, are finished. Adding the remaining time of the currently running
or scheduled steps to the left side of Constraint (\ref{eq:constr3_new}) can potentially
lead to triggering premature optimization rounds at $t+1$ when predecessor
steps of the next schedulable steps are not finished yet (and hence no new
steps can be scheduled).
\begin{eqnarray}
\tau_{t+1} + e_{i_{p}}
\leqslant DL_{i_{p}} + e_{i_{p}}^{p}
\label{eq:constr3_new}
\end{eqnarray}
Constraint (\ref{eq:constr4_new}) makes sure that the next optimization time period
$\tau_{t+1}$ is defined to be in the future. We use the value $\epsilon >
0$ to avoid optimization deadlocks arising from a too small value for
$\tau_{t+1}$.
The start of the next optimization period is directly calculated as a result of
the previous optimization rounds. Nevertheless, %as this optimization model only
%refers to one optimization round and only operates based on the current
%knowledge, 
the optimization can also be triggered by other events, like newly incoming
requests or finalized process steps and run earlier than the calculated
time $t+1$. In this case, a potentially still ongoing computation of an optimal task scheduling and resource allocation is canceled. 
\begin{eqnarray}
\tau_{t+1} \geqslant \tau_{t} + \epsilon
\label{eq:constr4_new}
\end{eqnarray}
Constraint (\ref{eq:constr5_new}) calculates the remaining enactment duration $e_{i_{p}}$ of a process instance. The remaining enactment duration does not
include the enactment and deployment time of the process steps that will be
scheduled for the current optimization period $\tau_{t}$, nor the remaining
enactment time of the currently already running steps. The overall remaining enactment duration $e_{i_{p}}$ of a process instance
$i_{p}$ is the sum of the enactment times of the different workflow patterns that the
process instance is composed of. We aggregate the upper-bound enactment times of
the different workflow patterns, following  the approach by \cite{jaeger:04}.
\begin{eqnarray}
e_{i_{p}} = e_{i_{p}}^{seq} + e_{i_{p}}^{L_{a}} + e_{i_{p}}^{L_{x}} +
e_{i_{p}}^{RL}
\label{eq:constr5_new}
\end{eqnarray}
The worst-case enactment times for sequences are defined in Constraint
(\ref{eq:constr6_new}), for AND-blocks in Constraint~(\ref{eq:constr7_new}), for XOR-blocks in
Constraint~(\ref{eq:constr8_new}), and for repeat loop blocks in Constraint~(\ref{eq:constr9_new}). For AND-blocks, the longest remaining enactment time of all paths within the block has to be considered. As we are using a worst-case analysis, we also consider the longest path for the calculation of the remaining enactment time of
XOR-blocks. For repeat loop blocks, the subpath enactment time is multiplied by
the maximum possible amount of repetitions $re$.

Constraints~(\ref{eq:constr6_new})--(\ref{eq:constr9_new}) consider the overall remaining
enactment time of all not yet running process steps for each process instance
$i_{p}$ as defined by the two helper variables $\hat{e}_{i_{p}}^{s}$ for
sequences and $\hat{e}_{i_{p}}^{l}$ for XOR- and AND-blocks in Constraints~(\ref{eq:constr11_new}) and (\ref{eq:constr12_new}), respectively. Since we perform a worst-case analysis,
both equations consider the worst-case VM startup time $\Delta$, i.e., including the time for
pulling the respective Docker image onto a running VM instance $\Delta_{st}$, and the
time for starting a new container instance $\Delta_{c_{st_{j}}}$ for each remaining process step
$j_{i_{p}}$.
If a process step $j_{i_{p}}^{*}$ is being scheduled in the current time period
$\tau_{t}$, Constraints~(\ref{eq:constr6_new})--(\ref{eq:constr9_new}) subtract the
worst-case enactment time of the scheduled steps $ex_{j_{i_{p}}^{*}}$ as defined in Constraint~(\ref{eq:constr10_new}), from the overall remaining enactment time.
\begin{eqnarray}
e_{i_{p}}^{seq} = \left\{ \begin{array}{ll}
\hat{e}_{i_{p}}^{s} - ex_{j_{i_{p}}^{*}}
&,\text{if\,\,} x_{(j_{i_{p}}^{*},k_{v},t)} = 1 \\
\hat{e}_{i_{p}}^{s} &, \text{otherwise}  \end{array} \right.
\label{eq:constr6_new}
\end{eqnarray}
\begin{eqnarray}
e_{i_{p}}^{L_{a}} = \left\{ \begin{array}{ll}
\displaystyle\max_{l \in La}(\hat{e}_{i_{p}}^{l} - ex_{j_{i_{p}}^{*}})
&,\text{if\,\,} x_{(j_{i_{p}}^{*},k_{v},t)} = 1 \\
\displaystyle\max_{l \in La}(\hat{e}_{i_{p}}^{l}) &, \text{otherwise}  \end{array} \right.
\label{eq:constr7_new}
\end{eqnarray}
\begin{eqnarray}
e_{i_{p}}^{L_{x}} = \left\{ \begin{array}{ll}
\displaystyle\max_{l \in Lx}(\hat{e}_{i_{p}}^{l} - ex_{j_{i_{p}}^{*}})
&,\text{if\,\,} x_{(j_{i_{p}}^{*},k_{v},t)} = 1 \\
\displaystyle\max_{l \in Lx}(\hat{e}_{i_{p}}^{l}) &, \text{otherwise}  \end{array} \right.
\label{eq:constr8_new}
\end{eqnarray}
\begin{eqnarray}
e_{i_{p}}^{RL} = \left\{ 
\begin{array}{ll}
	re \times \hat{e}_{i_{p}}^{s} - ex_{j_{i_{p}}^{*}}
	&, \text{if\,\,} x_{(j_{i_{p}}^{*},k_{v},t)} = 1 \\
	re \times \hat{e}_{i_{p}}^{s} &, \text{otherwise} 
	\end{array}
\right.
\label{eq:constr9_new}
\end{eqnarray}
\begin{eqnarray}
\begin{aligned}
ex_{j_{i_{p}}^{*}} = \sum_{\substack{v \in V}} \sum_{\substack{k_{v} \in K_{v}}}
 \big((e_{j_{i_{p}}^{*}} + \Delta_{c_{st_{j}}} + \Delta_{st_{j}} +
\Delta) \times x_{(j_{i_{p}}^{*},k_{v},t)}\big)
\label{eq:constr10_new}
\end{aligned}
\end{eqnarray}
\begin{eqnarray}
\hat{e}_{i_{p}}^{s} = \sum_{j_{i_{p}} \in J_{i_{p}}^{seq}}(e_{j_{i_{p}}} +
\Delta_{c_{st_{j}}} + \Delta_{st_{j}} + \Delta)
\label{eq:constr11_new}
\end{eqnarray}
\begin{eqnarray}
\hat{e}_{i_{p}}^{l} = \sum_{j_{i_{p}} \in J_{i_{p}}^{l}}(e_{j_{i_{p}}} +
\Delta_{c_{st_{j}}} + \Delta_{st_{j}} + \Delta)
\label{eq:constr12_new}
\end{eqnarray}
Constraints (\ref{eq:constr13_new}) and (\ref{eq:constr14_new}) consider for each
VM instance~$k_{v}$ the sum of CPU and RAM
resources required by all service invocations that either already run or are
scheduled to run in any container on the VM instance in $\tau_{t}$. These
resources need to be smaller or equal to the resource supply in terms of CPU
($s_{v}^{C}$) and RAM ($s_{v}^{R}$) that is offered by VM instances of
type $v$.
\begin{eqnarray}
\sum_{p \in P} \sum_{i_{p} \in I_{p}} \sum_{j_{i_{p}} \in (J_{i_{p}^{*}} \cup
	J_{i_{p}}^{run})} (r_{j_{i_{p}}}^{C} \times x_{(j_{i_{p}},k_{v},t)})
\leqslant
s_{v}^{C}
\label{eq:constr13_new}
\end{eqnarray}
\begin{eqnarray}
\sum_{p \in P} \sum_{i_{p} \in I_{p}} \sum_{j_{i_{p}} \in (J_{i_{p}^{*}} \cup
	J_{i_{p}}^{run})} (r_{j_{i_{p}}}^{R} \times x_{(j_{i_{p}},k_{v},t)})
\leqslant
s_{v}^{R}
\label{eq:constr14_new}
\end{eqnarray}
% Constraint 15 - 18 eliminated
Constraints (\ref{eq:constr19_new}) and (\ref{eq:constr20_new}) define for each VM instance $v~\in~V$, $k_{v}~\in~K_{v}$ the remaining free capacities in terms of CPU and RAM that are not
directly allocated to service instances. % (which are later packed into container instances).
As expressed by the variable $g_{(k_{v}, t)} \in \left\{0,1\right\}$, which is
further defined in Constraints (\ref{eq:constr21_new}) and (\ref{eq:constr22_new}), we only
consider VM instances that are either already running or are being leased
in $\tau_{t}$.
\begin{eqnarray}
g_{(k_{v}, t)} \times s_{v}^{C}  - \sum_{p \in P} \sum_{i_{p} \in I_{p}} \sum_{j_{i_{p}} \in (J_{i_{p}^{*}} \cup
	J_{i_{p}}^{run})}
 (r_{j_{i_{p}}}^{C} \times x_{(j_{i_{p}},k_{v},t)})
\leqslant
f_{k_{v}}^{C}
\label{eq:constr19_new}
\end{eqnarray}
\begin{eqnarray}
g_{(k_{v}, t)} \times s_{v}^{R} -  \sum_{p \in P} \sum_{i_{p} \in I_{p}} \sum_{j_{i_{p}} \in (J_{i_{p}^{*}} \cup
	J_{i_{p}}^{run})} (r_{j_{i_{p}}}^{R} \times x_{(j_{i_{p}},k_{v},t)})
\leqslant
f_{k_{v}}^{R}
\label{eq:constr20_new}
\end{eqnarray}
Constraints (\ref{eq:constr21_new}) and (\ref{eq:constr22_new}) define the value of
the helper variable $g_{(k_{v}, t)} \in \left\{0,1\right\}$ which takes the
value of 1 if a VM instance is either already running ($\beta_{k_{v}} =
1$) or being leased for at least one BTU ($y_{(k_{v},t)} \geq 1$) in
$\tau_{t}$. $g_{(k_{v}, t)}$ is restricted to be a boolean variable in
Constraint~(\ref{eq:constr45_new}).
\begin{eqnarray}
g_{(k_{v}, t)}
\leqslant
\beta_{(k_{v}, t)} + y_{(k_{v}, t)}
\label{eq:constr21_new}
\end{eqnarray}
\begin{eqnarray}
\beta_{(k_{v}, t)} + y_{(k_{v}, t)}
\leqslant
BTU^{max} \times g_{(k_{v}, t)}
\label{eq:constr22_new}
\end{eqnarray}
Constraint (\ref{eq:constr23_new}) ensures that for all VM instances $v \in
V, k_{v} \in K_{v}$ a VM instance $k_{v}$ will be leased or is running
if service invocations are to be placed %(within later defined containers) 
on it.
Therefore, we calculate the number of process steps that are scheduled or already running on $k_{v}$ and make use of
our helper variable $g_{(k_{v}, t)}$ as defined in Constraints (\ref{eq:constr21_new})
and~(\ref{eq:constr22_new}), as well as a helper variable $(M \times N)$ that presents an
arbitrarily large number, e.g., 100 $\times$ 10,000, as an upper bound of possible
container deployments per VM instance times possible parallel service invocations per container instance. Notably, this upper bound does \emph{not} imply that an according number of service invocations are actually carried out. Instead, the goal of this variable is to avoid that the optimization gets stuck in a loop.
\begin{eqnarray}
\sum_{p \in P} \sum_{i_{p} \in I_{p}} \sum_{j_{i_{p}} \in (J_{i_{p}}^{*} \bigcup
	J_{i_{p}}^{run})} x_{(j_{i_{p}}, k_{v}, t)}
\leqslant
g_{(k_{v}, t)} \times (M \times N)
\label{eq:constr23_new}
\end{eqnarray}
% Constraint 24 - 29 eliminated
Constraint (\ref{eq:constr30_new}) makes sure that for all VM instances $v \in
V$, $k_{v} \in K_{v}$ and all running or schedulable process steps $j_{i_{p}} \in  (J_{i_{p}}^{*} \bigcup
J_{i_{p}}^{run})$, the scheduled service invocations $j_{i_{p}}$ will not be moved to other VM instances.
First, we consider the remaining enactment time of a
process step $j_{i_{p}}$ for each process step scheduled on a
VM instance ($x_{(j_{i_{p}}, k_{v}, t)} = 1$). In case that the process
step $j_{i_{p}}$ is not running yet, we also have to consider the time to pull a new
image $\Delta_{st_{j}}$ and to start the needed container $\Delta_{c_{st_{j}}}$, if the image does not exist and the
container is not running on the scheduled VM instance ($z_{(st_{j}, k_{v},
	t)} = 0$). Furthermore, if the required VM instance~$k_{v}$ is not up
and running yet ($\beta_{(k_{v}, t)} = 0$), we also need to add the time $\Delta$ for
starting the new VM instance $k_{v}$ to our calculation of the remaining
enactment time. We demand the calculated remaining enactment time to be smaller or equal to the remaining
leasing duration $d_{(k_{v}, t)}$ of the allocated VM instance $k_{v}$.
If the currently remaining leasing duration is smaller, the corresponding
VM instance~$k_{v}$ needs to be leased for $y_{(k_{v}, t)} \in \left\{\mathbb{N}_{0}^{+}\right\}$ additional
BTUs in the time period $\tau_{t}$ to satisfy the constraint.
\begin{eqnarray}
\begin{split}
\big(e_{j_{i_{p}}}
+ (\Delta_{c_{st_{j}}} + \Delta_{st_{j}}) \times (1 - z_{(st_{j}, k_{v}, t)})
+ \Delta \times (1 - \beta_{(k_{v}, t)})\big)\\
\times x_{(j_{i_{p}}, k_{v}, t)} 
\leqslant
d_{(k_{v}, t)} \times \beta_{(k_{v}, t)} + BTU_{k_{v}} \times y_{(k_{v}, t)}
\end{split}
\label{eq:constr30_new}
\end{eqnarray}
% Constraint 31 - 33 eliminated
Similar to Constraint (\ref{eq:constr30_new}), Constraint (\ref{eq:constr34_new})
together with Constraint (\ref{eq:constr38_new}) ensures
that all service invocations that are already running in a container on a certain
VM instance at the start of $\tau_{t}$, will not be
migrated to another VM instance (or container) during their invocation. We
explicitly reference the already assigned
VM instance $k_{v}$ by the additional indices in Constraint~(\ref{eq:constr34_new}) and do not need to consider additional times for pulling
container images or starting VM instances.
\begin{eqnarray}
e_{j_{i_{p}}}^{run_{k_{v}}}
\leqslant
d_{(k_{v}, t)} \times \beta_{(k_{v}, t)} + BTU_{k_{v}} \times y_{(k_{v}, t)}
\label{eq:constr34_new}
\end{eqnarray}
Constraint~(\ref{eq:constr35_new}) defines the variable $\gamma_{(v, t)}$ for all
VM types $v \in V$. $\gamma_{(v, t)}$ indicates the total number of
BTUs to lease VM instances of type $v$, meaning it includes the
decisions which VM instances $k_{v}$ to lease and for how long. To define $\gamma_{(v, t)}$, we build the sum over
all leased BTUs for each VM instance of type $v$.
\begin{eqnarray}
\sum_{k_{v} \in K_{v}} y_{(k_{v}, t)}
\leqslant
\gamma_{(v, t)}
\label{eq:constr35_new}
\end{eqnarray}
Constraint~(\ref{eq:constr36_new}) demands for all next schedulable process steps
$j_{i_{p}} \in J_{i_{p}}^{*}$ that each service invocation is scheduled on only one VM instance
(and later enacted within only one container instance) or postponed for a later optimization period.
\begin{eqnarray}
\sum_{\substack{v \in V}} \sum_{\substack{k_{v} \in K_{v}}} \sum_{\substack{j_{i_{p}} \in J_{i_{p}}^{*}}} x_{(j_{i_{p}}, k_{v}, t)}
\leqslant
1
\label{eq:constr36_new}
\end{eqnarray}
% Constraint 37 eliminated
Constraint~(\ref{eq:constr38_new}) makes sure that all steps currently running on a VM instance $k_{v}$
remain on the same VM instance. This is achieved by setting the decision variable $x_{(j_{i_{p}}, k_{v},
	t)}$ for each already running service invocation $j_{i_{p}}^{run}$ on a VM instance $k_{v}$ to 1.
This constraint corresponds to inheriting the scheduling decision already
made in a previous period.
\begin{eqnarray}
x_{(j_{i_{p}}^{run}, k_{v}, t)} = 1
\label{eq:constr38_new}
\end{eqnarray}
% Constraint 39 eliminated
The remaining Constraints~(\ref{eq:constr40_new})--(\ref{eq:constr45_new}) restrict the values particular variables can adopt. 
Constraint~(\ref{eq:constr40_new}) restricts the decision variable  $x_{(j_{i_{p}},
	k_{v}, t)}$ that decides whether or not to schedule a service invocation
$j_{i_{p}}$ on a certain VM instance~$k_{v}$ for all $p \in P, i_{p} \in I_{p},
j_{i_{p}} \in J_{i_{p}}^{*}$ to be a boolean.
\begin{eqnarray}
x_{(j_{i_{p}}, k_{v}, t)} \in \left\{0,1\right\}
\label{eq:constr40_new}
\end{eqnarray}
% Constraint 41 eliminated
Constraint~(\ref{eq:constr42_new}) restricts the decision variable $y_{(k_{v}, t)}$
that decides for how many additional BTUs to lease a VM instance
$k_{v}$ at time $t$ for all $v \in V, k_{v} \in K_{v}$ to be a positive integer
$\geq 0$. Constraint~(\ref{eq:constr43_new}) does the same for  $\gamma_{(v,t)}$ that decides for how many total BTUs to lease any VM instances
\begin{eqnarray}
y_{(k_{v}, t)} \in \mathbb{N}_{0}^{+}
\label{eq:constr42_new}
\end{eqnarray}
\begin{eqnarray}
\gamma_{(v,t)} \in \mathbb{N}_{0}^{+}
\label{eq:constr43_new}
\end{eqnarray}
Constraint~(\ref{eq:constr44_new}) restricts the variable that indicates the penalty
enactment time $e_{i_{p}}^{p}$ to be a positive real number.
\begin{eqnarray}
e_{i_{p}}^{p} \in \mathbb{R}^{+}
\label{eq:constr44_new}
\end{eqnarray}
Constraint (\ref{eq:constr45_new}) restricts the variable $g_{(k_{v}, t)}$ to be a
boolean. The variable has been defined in Constraints~(\ref{eq:constr21_new}) and
(\ref{eq:constr22_new}) and takes the value of 1 if a VM instance is either
already running ($\beta_{k_{v}} = 1$) or being leased for at least one BTU
($y_{(k_{v},t)} \geq 1$) in $\tau_{t}$.
\begin{eqnarray}
g_{(k_{v}, t)} \in \left\{0,1\right\}
\label{eq:constr45_new}
\end{eqnarray}

% Constraint 46 - 47 eliminated

\subsection{Transformation Step}
\label{sub:transformation}
Counterintuitively, the problem discussed so far partly disregards the container allocation: it schedules process steps directly on VM instances, while considering the current information about deployed containers. More formally, the output of the model is a task scheduling and resource allocation decision stating for all VM instances $v \in V, k_{v} \in K_{v}$ what VM instances to lease for how many BTUs $y_{(k_{v},t)}$ and for all next schedulable or
running process steps $j_{i_{p} \in J_{i_{p}}^{*} \bigcup J_{i_{p}}^{run}}$ on which VM instance $v \in V, k_{v} \in K_{v}$ to schedule the service invocations $x_{(j_{i_{p}}, k_{v}, t)}$. 

This means that the actual container instances are not yet part of the scheduling plan after the optimization. While this information could also be derived from an extended version of the presented MILP problem, this would make the FFSIPP even more complex. 

We opted against this for several reasons: Most importantly, the extended optimization model would have to handle a very large number of decision variables: First, a MILP solver has to decide for each VM instance in the set of leasable instances whether or not to lease the VM instance. Second, the solver would need to decide for each container instance whether or not to deploy a container with a certain configuration on a certain VM instance. Third, the solver would need to decide on which deployed container instance to schedule the service requests. When doing so, the computational complexity grows exponentially with the number of incoming process requests and also with the number of VM instances and container instances. 

Related work in the field of MILP-based optimization problems has shown that solutions struggle with increasing problem sizes~\citep{mann:15}. In order to avoid this, we introduce a so-called transformation step. 
This step wraps each scheduled process step of a certain service type in a  container instance of the same service type, which is to be deployed on the respective VM as per the optimization (see Section~\ref{subsub:containers}). This guarantees the exact resources, as needed by the scheduled service invocations on the VM. Generally speaking, the transformation step described in Algorithm~\ref{alg:transformation} decreases the problem size by avoiding that a MILP solver has to decide which containers with which configurations to deploy on the VM instances.

To transform the result of the optimization model to a result which takes into account container instances, Algorithm~\ref{alg:transformation} iterates over all VM instances $k_{v}$ of the scheduling plan (line~1) and extracts all process steps that are scheduled for enactment on a VM instance (line~2). For all process steps $j_{i_{p}}$ of the same service type $st$ (lines 3-5), a container instance $c_{st}$ is defined (line 7) with a configuration that makes sure that the containers minimum resource supply \emph{coSize} is set to be exactly as large as the sum of scheduled service invocations of type $st$ on $k_{v}$ (line 13). We can then exchange the variable $x_{(j_{i_{p}}, k_{v}, t)}$ from the MILP solution for the introduced variables $x_{(j_{i_{p}}, c_{st}, t)}$ and $a_{(c_{st}, k_{v} t)}$ (indicating if the container $c$ with the service type $st$ should be deployed on VM $k_{v}$ in the time period $\tau_{t}$), which we set to 1 (lines 11, 14). This means, our scheduling plan schedules all service invocations $j_{i_{p}}$ of service type $st$ that were scheduled on $k_{v}$ on the newly defined container instance $c_{st}$, while the container instance
$c_{st}$ is scheduled on $k_{v}$. 

\begin{algorithm}[t]
	\caption{Transformation}
	\label{alg:transformation}
	\small
	\begin{algorithmic}[1]
	\ForAll{$k_{v}$ in the scheduling plan}
		\State $VMServiceMap \gets$ new $Map$($st$, $List$($j_{i_{p}}$));
		\ForAll{$j_{i_{p}}$ $where$ $x_{(j_{i_{p}}, k_{v}, t)} = 1$}
			\State $VMServiceMap \gets$ $(st_{j}, j_{i_{p}}) $;
		\EndFor
		\ForAll{$st$ in $VMServiceMap.getKeys()$}
			 \State $c_{st} \gets $ new $Container(st)$;
			 \State $coSize \gets 0$;
			 \ForAll{$j_{i_{p}}$ in $VMServiceMap.getValues(st)$}
				\State $coSize \gets (coSize +j_{i_{p}}.getResSize())$;
				\State $x_{(j_{i_{p}}, c_{st}, t)} \gets 1$;
			\EndFor
			\State $c_{st}.setSize(coSize)$;
			\State $a_{(c_{st}, k_{v} t)} \gets 1$;
		\EndFor
	\EndFor
\end{algorithmic}
\end{algorithm}

\subsection{Enacting the Transformed Scheduling Plan}
When enacting the transformed scheduling plan, the main tasks of the
eBPMS are to lease or release VM instances, to deploy, stop, or
adjust containers, and to place the service invocations. For each task, different
rules have to be considered:

%\paragraph{Enacting the VM Plan}
First, whenever the decision variable $y_{(k_{v}, t)}$ takes a value $>~0$, the
corresponding VM instance $k_{v}$ needs to be leased or the lease of the
VM instance needs to be extended by more BTUs. If during an
optimization round no containers are scheduled on a VM instance $k_{v}$,
the instance will still continue to be running until the end of its
leasing duration.

%\subsubsection{Enacting the Container Plan}
%Container instances are more flexible than the VM instances. 
Second, whenever the reasoner delivers the solution $a_{(c_{st}, k_{v}, t)} = 1$ for a
container instance $c_{st}$, either the deployment of the instance is initiated
by the eBPMS if a container of type $st$ is not running on $k_{v}$ at time~$t$, or a currently running container of type $st$ on $k_{v}$ is potentially
resized to match the defined container configuration. If a container instance of
service type $st$ on a VM instance $k_{v}$ is currently running during 
$\tau_{t}$, but the scheduling plan does not define a corresponding container
scheduling of the form $a_{(c_{st}, k_{v}, t)} = 1$, the running container is
immediately shut down to free up resources for other containers.

%\subsubsection{Enacting the Service Invocation Plan}
Third, all scheduled service invocations $x_{(j_{i_{p}}, c_{st}, t)} = 1$ that are not
yet running are invoked by the eBPMS on the defined container $c_{st}$. As
the scheduling plan also contains running steps, the enactment of those service
invocations that have already been scheduled for the first time in a previous
optimization round remains unchanged.
\section{Evaluation}
\label{sec:evaluation}
\begin{table}[h]
	\centering
	\caption{Evaluation Process Models}
	\label{tab:processModels}
	\footnotesize
	\begin{tabular}{c|c|c|c|c}
		\textbf{Process No.} & \textbf{$|$Steps$|$} & \textbf{$|$XOR$|$} & \textbf{$|$AND$|$} & \textbf{$|$Loops$|$} \\
		\hline 
		1 & 3 & 0 & 0 & 0\\
		2 & 2 & 1 & 0 & 0\\
		3 & 3 & 0 & 1 & 0\\
		4 & 8 & 0 & 2 & 0\\
		5 & 3 & 1 & 0 & 0\\
		6 & 9 & 1 & 1 & 0\\
		7 & 8 & 0 & 0 & 0\\
		8 & 3 & 0 & 1 & 0\\
		9 & 4 & 0 & 1 & 1\\
		10 & 20 & 0 & 4 & 0\\
	\end{tabular}
\end{table}
In this section, we conduct a quantitative analysis of the implementation of the FFSIPP as presented in Section~\ref{sec:problem}. As testbed, we use an extended version of the eBPMS ViePEP~\citep{hoenisch:16,WHS+19}, which also provides the performance metrics needed in our optimizations. For the predictions of CPU demands, we apply the values mentioned in Tables~\ref{tab:evalServicesResourceIntensive} and~\ref{tab:evalServicesLessResourceIntensive}. However, we could also apply a prediction approach for this, e.g.,~\cite{BSH16}. 

All evaluation runs have been executed on a MacBook Pro with an \emph{i5} quad core CPU with 2.50 GHz, 8 GB of RAM, running Ubuntu 16.04 with Linux kernel 4.8.0. As MILP solver, IBM CPLEX is used. Internally, CPLEX makes use of different (simplex) algorithms.

%MacBook Pro (Retina, 13-inch, Late 2012)
%Processor: 2,5 GHz Intel Core i5
%Memory: 8 GB 1600 MHz DDR3
%Graphics: Intel HD Graphics 4000 1536 MB

\subsection{Evaluation Setup}
\label{sub:evasetup}
\subsubsection{Process Models}
\label{subsub:processmodels}
We perform our evaluation using a subset of the SAP reference model~\citep{curran:97}, which has been used for
multiple scientific papers, e.g., \citep{mendling:08}, and provides a
solid foundation for our evaluations. From the around 600 process models in the reference model, we
select ten models with different process patterns and varying levels of
complexity, including sequences, XOR-blocks, AND-blocks, and repeat loops. All XOR- and
AND-blocks start with a split and end with a merge pattern. The merge for
AND-blocks is blocking, while the merge for XOR-blocks continues the process
enactment as soon as the last process step of one optional branch is finished.
Our chosen process models range from simple sequences to complex process models (see Table~\ref{tab:processModels}). %Since the work at hand is about resource allocation and task scheduling on cloud-based computational resources, we generally only consider software-based services for process steps.% and no human-provided services. 

For the single services providing the needed software functionalities of process steps, we take into account two different settings in order to evaluate the performance of our approach: In Section~\ref{sub:intensive}, we use resource-intensive services for the single process steps. In Section~\ref{subsub:lessresource}, we apply less resource-intensive services in order to evaluate whether this aspect leads to significant differences in the evaluation results. We assume that the service makespans are rather long, i.e., that the services cover long-running computational tasks.

\subsubsection{Applied SLAs}
\label{subsub:slas}
%For the enactment of each process instance, different deadline SLAs may be defined on the process level with regard to the complete process enactment.

We evaluate our approach using two different process deadline scenarios. In the
first scenario, we use strict deadlines and demand the latest point in time
when the enactment of process instances has to be finished to be 1.5 times
the process model's average makespan from the point in time when the request is
sent. The second scenario allows more lenient SLA values and we demand
2.5 times of the process model's average makespan as the maximum enactment
duration. These values were chosen based on preliminary evaluation runs and to account for the
startup time of VM instances and the time to pull Docker images and deploy
container instances.

For calculating penalty cost, we apply a linear cost model which assigns one unit
of penalty cost per 10\% of time units of delay.

\subsubsection{Process Request Arrival Patterns}
\label{subsub:arrivalpatterns}
Our experimental design accounts for two different process request arrival
patterns. In one scenario, we design a \textit{constant} arrival pattern.
Specifically, we choose to request two process instances every two minutes,
alternating process instance requests for process models 1 to 5 and 6 to 10
until 50 process instance requests have been sent.
In a second scenario, we follow a \textit{pyramid}-like function for designing our arrival
pattern. We send a total of 100 randomly shuffled process instance requests in
different batch sizes, ranging from one to five instances at a time in an interval
of one minute. Equation~(\ref{eq:pyr}) represents the pyramid pattern, $n$
representing the time in minutes since the start of the experiment, and $a$ representing the
number of new process requests.

\begin{eqnarray}
f(n) = a
\begin{cases}
1 & if 0 \leq n \leq 4 \\
\lceil(n+1)/4\rceil & if 5 \leq n \leq 17 \\
0 & if 18 \leq n \leq 19 \\
1 & if 20 \leq n \leq 35 \\
\lceil(n-9)/20\rceil & if 36 \leq n \leq 51 \\
\end{cases}
\label{eq:pyr}
\end{eqnarray}
These two different arrival patterns have been chosen in order to evaluate the presented approach in a rather stable environment, with process requests coming in constantly, and a more volatile environment, where the requests do not follow such a constant pattern. This may lead to different results, since our optimization model does not regard process request arrivals. However, it might be interesting to integrate such knowledge in the future (see Section~\ref{sec:conclusion}). %Similar patterns have been used in our former work on evaluating task scheduling and resource allocation approaches for elastic processes~\citep{WHS+19,hoenisch:16}.

Notably, we have also tested other arrival patterns and settings in preliminary evaluation runs. The presented settings have been chosen, since they are representative for the observations we have made in our preliminary tests. 

\subsection{Test Environment}
\label{sub:test}

%A rather large amount of research on scheduling and resource allocation problems for the cloud uses self-constructed simulation environments, but also off-the-shelf simulation frameworks like, e.g., CloudSim\footnote{\url{http://www.cloudbus.org/cloudsim/}} for evaluation purposes~\citep{mann:15}. To the best of our knowledge, there is no widely-accepted simulation framework for elastic processes. 
For allowing %large and 
reproducible test scenarios over multiple hours and
with different pricing models, we perform the optimization runs using the
simulation mode of ViePEP. This means that a complete process landscape is modeled and requested using the eBPMS, and the optimization model is able to operate with these values. However, the actual leasing of resources is not activated. Instead, the utilization of VMs is calculated based on the characteristics of the evaluation services (see Sections~\ref{sub:intensive} and \ref{subsub:lessresource}).

Our system considers seven VM types with the characteristics listed in Table~\ref{tab:VMsSimulation}. As it can be seen, we allow both privately hosted and publicly hosted VMs. Regarding the pricing model, we assume that leasing a VM instance with $x$ cores is cheaper than leasing two VM instances with $x/2$ cores, i.e., the marginal costs are decreasing. We apply a BTU of five minutes for both private and public cloud resources. Also, we assume that VMs from the private cloud are cheaper to lease than equally sized VMs from the public cloud. 

\begin{table}[t]
	\caption{Evaluation VM Types}
	\label{tab:VMsSimulation}
	\centering
	\footnotesize
	\begin{tabular}{c|c|c|c}
		\textbf{VM Type Name} &  \textbf{Provider} & \textbf{CPU Cores} &
		\textbf{Cost per BTU} \\
		\hline 1
		& Private
		& 1
		& 10
		\\
		2
		& Private
		& 2
		& 18
		\\
		3
		& Private
		& 4
		& 30
		\\
		4
		& Public A
		& 1
		& 15
		\\
		5
		& Public A
		& 2
		& 25
		\\
		6
		& Public A
		& 4
		& 35
		\\
		7
		& Public A
		& 8
		& 50
		\\
	\end{tabular}
\end{table}

As it can be seen in the table, we do not further regard RAM utilization, even though RAM is explicitly foreseen in the problem formulation presented in Section~\ref{sec:problem}. The reason for this is that preliminary results have shown that since both RAM and CPU are additive resources, the RAM utilization does not impact the overall cost too much. So, in order to keep the evaluation section concise, we opted to not further discuss it here. 

In the evaluation, we assume a VM startup time of 60~seconds, a time to pull Docker images from the central registry onto a VM instance of
30 seconds, and a time to start a new Docker container and the contained service from an existing image of two seconds. These values have been chosen based on recommendations in the related work~\citep{xavier:16,lingayat18,velp20}.

\subsection{Metrics}
\label{subsec:eval:metrics}
We assess the outcome of our experiments by using different metrics: Minimizing the \emph{total cost} is the core goal of FFSIPP. Accordingly, we measure the cost arising for enacting a complete business process landscape over time. These cost include the VM leasing and penalty cost over all process enactments during an experiment's runtime.

In addition, we measure two secondary metrics in order to assess if FFSIPP also contributes to achieving them: \emph{SLA adherence} calculates the percentage of process requests which meet their SLAs, i.e., the process instances that finish their enactment before the defined deadline. As described in Section~\ref{sub:objective}, SLA adherence is not a hard constraint in our optimization model, but is indirectly achieved by taking into account that penalty cost accrue if an SLA is violated. 

In addition, we also measure the \emph{makespan}, i.e., the total duration for enacting all incoming process requests in a particular experiment, from the first-received request to the last finished process step. Since the arrival of a process request is not needed in our system model, this value is monitored by our evaluation environment, i.e., ViePEP. %Again, the makespan is not an explicit goal of our optimization, but it could be decreased by the better utilization of available computational resources. 

Experiments are repeated three times in order to take into account that the duration of a service invocation may have some spread, as described in Section~\ref{sub:intensive}. Accordingly, we calculate the standard deviations $\sigma$ for all metrics.

\subsection{Baseline}
\label{sub:baseline}
As the baseline for the evaluation of FFSIPP, we apply a slightly modified state-of-the-art approach for VM-based resource allocation and task scheduling by Hoenisch et al.~\citep{hoenisch:16}, called Service Instance Placement Problem (SIPP). The SIPP approach uses an optimization model that directly schedules process requests on VM instances, while each leased VM instance is only able to enact process requests of one specific service type. Containers are not regarded.

The modifications of the originally presented baseline are made with regard to the utilized time constraints which we define according to the time constraints also used by our approach as presented in Constraints (\ref{eq:constr2_new}) and (\ref{eq:constr3_new}).
Furthermore, we allow the change of the offered service type of a VM instance during an uninterrupted leasing duration
whenever no more service requests are being enacted by a VM instance.
When changing the offered service type, a service deployment time of another 30 seconds is considered and again only one service type is offered at any point in time by a VM instance in the baseline.

\subsection{Results}
\subsubsection{Resource-intensive Services}
\label{sub:intensive}

\begin{table}[h]
	\centering
	\caption{Evaluation Services -- Resource-Intensive}
	\label{tab:evalServicesResourceIntensive}
		\footnotesize
	\begin{tabular}{c|c|c}

		& \textbf{CPU Load} &
		\textbf{Service Makespan} \\
		\textbf{Service Type Name} & \textbf{in \% ($\mu_{cpu}$)} & \textbf{in sec. ($\mu_{dur}$)} \\
		\hline A
		& 45
		& 40
		\\
		B
		& 75
		& 80
		\\
		C
		& 75
		& 120
		\\
		D
		& 100
		& 40
		\\
		E
		& 120
		& 100
		\\
		F
		& 125
		& 20
		\\
		G
		& 150
		& 40
		\\
		H
		& 175
		& 20
		\\
		I
		& 250
		& 60
		\\
		J
		& 333
		& 30
		\\
	\end{tabular}
\end{table}
As mentioned in Section~\ref{subsub:processmodels}, we apply two different types of services in our evaluation. First, we discuss rather resource-intensive services, while services with smaller resource demands are covered in Section~\ref{subsub:lessresource}.

The characteristics of the services for the scenario at hand are presented in Table~\ref{tab:evalServicesResourceIntensive}. 
The values in Table \ref{tab:evalServicesResourceIntensive} represent mean values of general normal
distributions with $\sigma_{cpu} = \mu_{cpu}/10$ and $\sigma_{dur} =
\mu_{dur}/10$. This is done in order to reflect that in real-world settings, services will not always have the same makespan, e.g., since the base loads of a container or VM may differ. 
A services' CPU load refers to the required amount of CPU resources in percent
of a single core, meaning all services with a CPU requirement of more
than 100\%, cannot be scheduled on a single-core VM and the corresponding
containers will need to be placed on a multi-core VM instance. We assume that services are fully parallelizable.

%\subsubsection{Results and Discussion}
%\label{subsec:eval:result}

\paragraph{Constant Arrival of Resource-Intensive Processes}
\begin{figure}[h]
	\centering
	\subfloat[][Constant Arrival, Strict
	SLA]{\includegraphics[width=.35\textwidth]{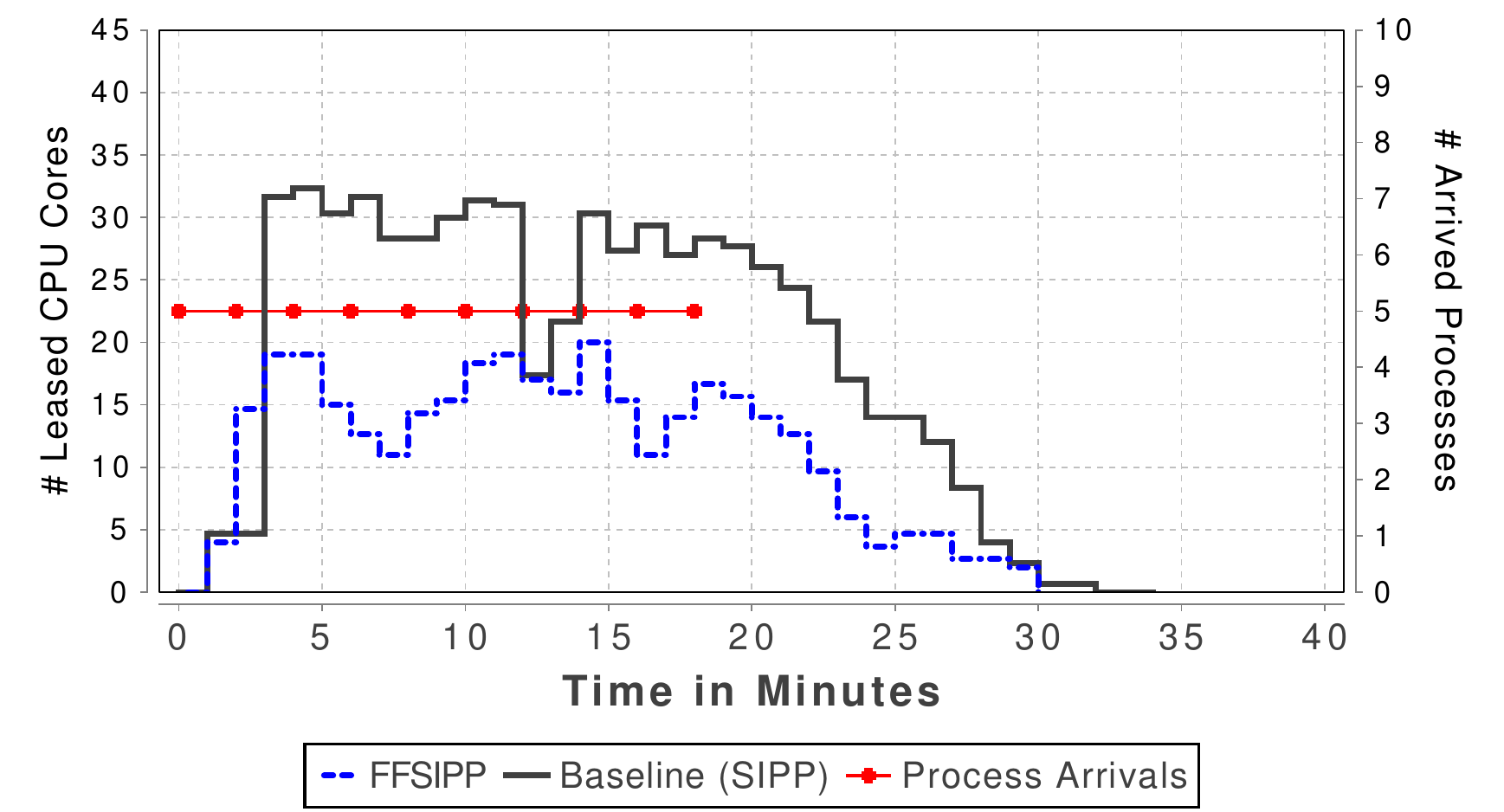}\label{subfig:constant15_intense:strict}}
	\\
	\subfloat[][Constant Arrival, Lenient
	SLA]{\includegraphics[width=.35\textwidth]{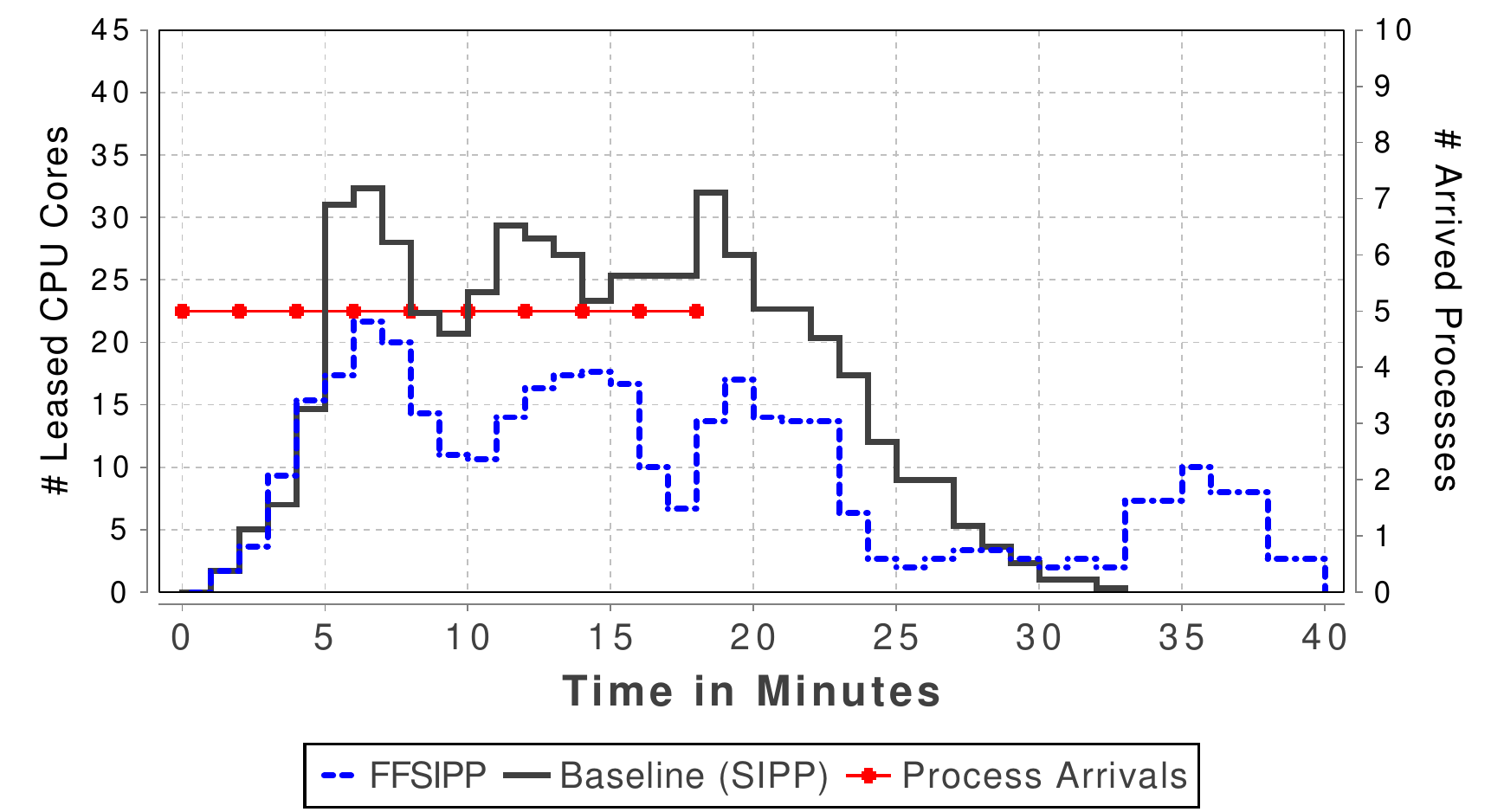}\label{subfig:constant15_intense:lenient}}
	%	\caption{Evaluation Results -- Constant Arrival of Resource-Intensive Processes}
	\caption{Evaluation Results -- Constant Arrival (Resource-Intensive Services)}
		\label{fig:constant15}
\end{figure}
\begin{figure}[t]
	\centering
	\label{fig:constant15_pyramid}
	\subfloat[][Pyramid Arrival, Strict
	SLA]{\includegraphics[width=0.35\textwidth]{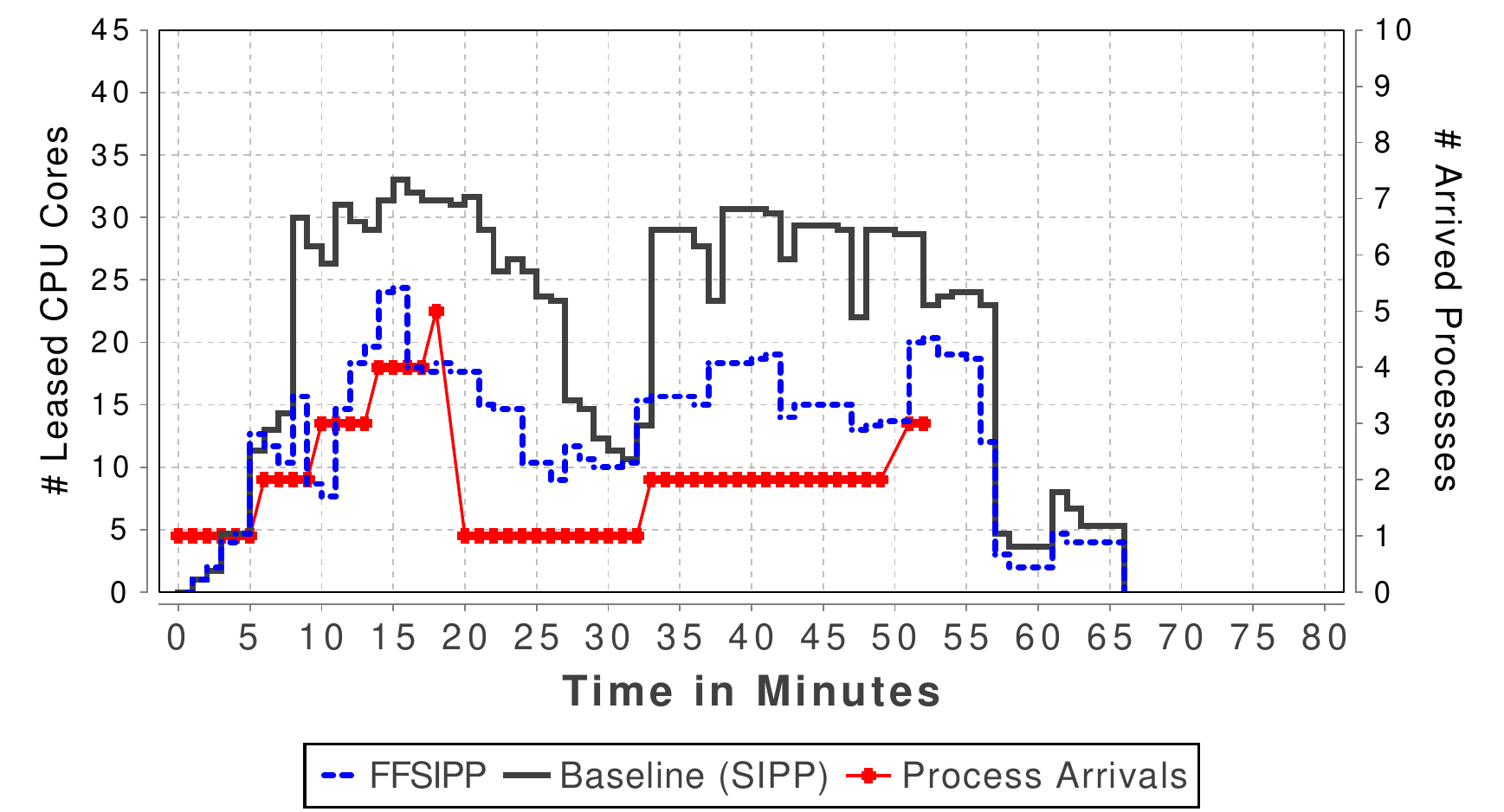}\label{subfig:pyramid15_intense:strict}}
	\\
	\subfloat[][Pyramid Arrival, Lenient
	SLA]{\includegraphics[width=0.35\textwidth]{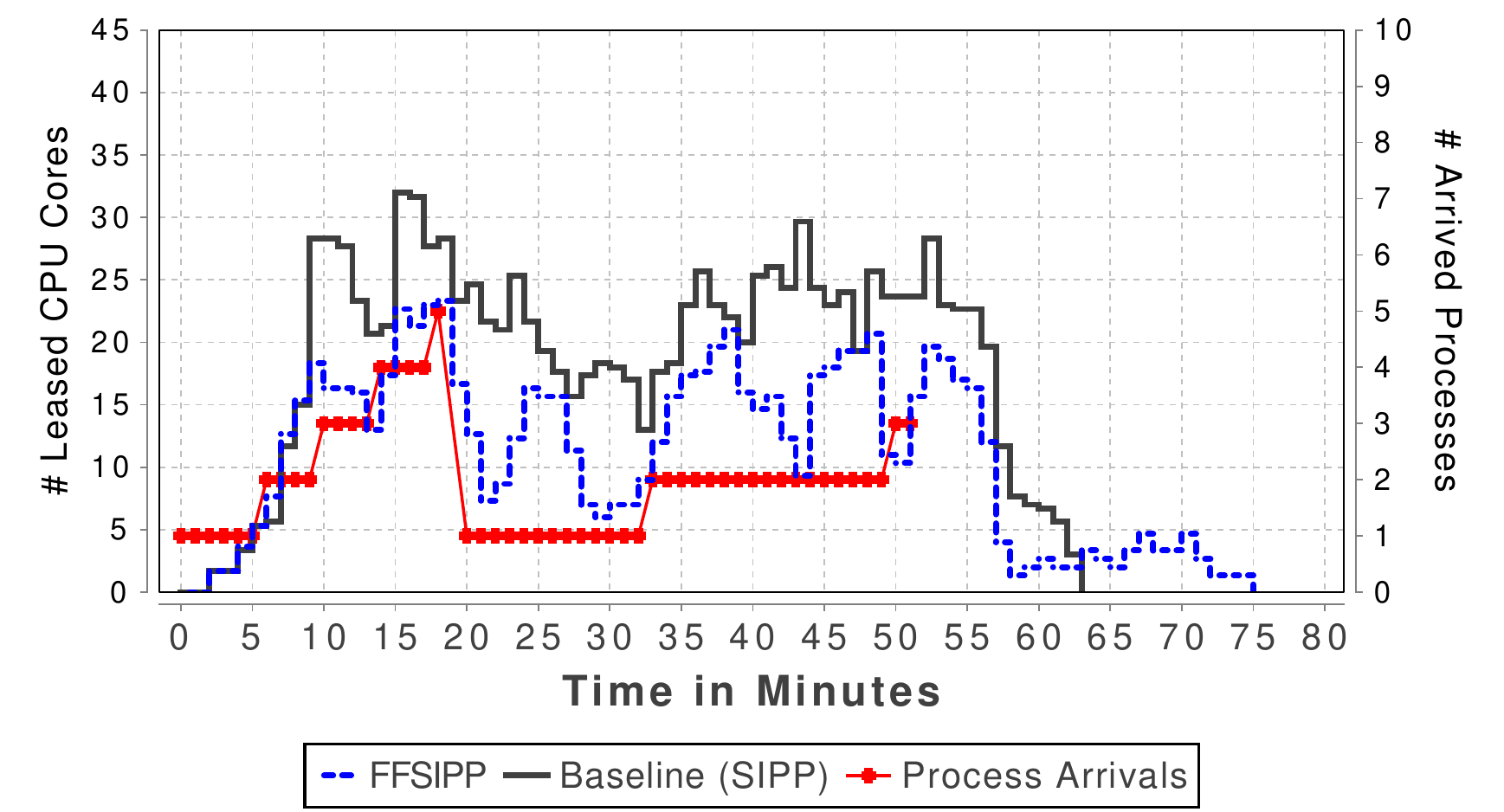}\label{subfig:pyramid15_intense:lenient}}
	\caption{Evaluation Results -- Pyramid Arrival (Resource-Intensive Services)}
\end{figure}

First, we compare the results for resource-intensive services for the constant arrival pattern. Figure \ref{subfig:constant15_intense:strict} shows the results for process requests with a strict deadline, while Figure \ref{subfig:constant15_intense:lenient} depicts the results for the lenient SLA.
The charts plot the leased CPU cores over time and also show the corresponding process request arrival patterns.
The horizontal axes show the time in minutes. The left vertical axes display the number of leased CPU cores
while the right vertical axes represent the number of parallel process requests. 
An overview of the results is shown in Table~\ref{tab:evaluationReducedToBaseline:resourceIntensive:constant}. The table presents the numerical average values of the evaluation runs and the standard deviations. It is important to distinguish between the total makespan in minutes as discussed in Table~\ref{tab:evaluationReducedToBaseline:resourceIntensive:constant}, which
refers to the time for enacting all incoming process requests, and the times shown in Figure~\ref{fig:constant15} which
refer to time slots when VM instances (cores) were leased.

As can be seen from Table~\ref{tab:evaluationReducedToBaseline:resourceIntensive:constant}, our optimization approach yields a high SLA adherence of 98.67\% with a very small
standard deviation of only 1.15 percentage points (pp) and only 1 unit of penalty cost for process requests with strict enactment deadlines.
Compared to the baseline which only yields an SLA adherence of 86.00\% with a standard deviation of 2.00 pp, and on
average 15 units of penalty cost, the FFSIPP approach is significantly improving the detection of potential SLA violations and also able to rapidly calculate a good scheduling strategy.
Furthermore, regarding the leasing cost, our approach leads to only about half the cost of the baseline
with also a lower associated standard deviation.

When comparing the result for more lenient deadlines, we can observe a much better SLA adherence of 98.67\% for the
SIPP baseline (as compared to strict deadlines), but again a better SLA adherence of 100.00\% for all performed runs using the FFSIPP 
approach. This means our approach is able to finish all incoming process requests within the defined deadline constraints for
all evaluation runs.
The leasing cost  are again nearly half as high compared to those of the baseline. The low leasing cost of our
approach are a result of the near-optimal resource utilization that is realized through the dynamic assignment and reassignment
of Docker containers for the enactment of different service types that only reserve as much space from the underlying VM
instances as they actually require.

Notably, the enactment of process requests using our approach with lenient process deadlines leads to a longer total
makespan and lower total leasing cost compared to the enactment of process requests with a strict deadline.
When looking at the number of leased resources in Figure \ref{subfig:constant15_intense:lenient}, we can clearly observe
that our approach leases a smaller amount of computational resources in terms of CPU cores than the SIPP baseline at most points in time, and that
our approach is able to postpone the enactment of process steps with a distant deadline for as long as possible, if that is beneficial in terms of cost. This explains why after 33 minutes the number of leased resources increases after it had already declined. 
The baseline approach does not make use of the more distant deadline constraints in a similarly efficient way, and also does not utilize the existing resources to the same degree as it is possible with containers. Here, the shortcoming that the baseline assumes that only one service can be deployed on one VM at the same time, is one major cost driver. 
The reason
why the makespan for the baseline is similar for both the strict and the lenient process deadlines is that there are already
sufficient computational resources leased at earlier points in time. These resources were not fully utilized and
were able to perform the enactment of process steps ahead of their scheduling deadlines.\\

\begin{table*}[t]
		\caption{Evaluation Results -- Resource-Intensive Services}
	\label{tab:evaluationReducedToBaseline:resourceIntensive:constant}
	\centering
\begin{adjustbox}{width=0.85\linewidth,center}
\begin{tabular}{l|c|c|c|c|c|c|c|c}
	\textbf{Arrival Pattern} &  \multicolumn{4}{c|}{\textbf{Constant Arrival}}
	& \multicolumn{4}{c}{\textbf{Pyramid Arrival}} 
	\\
	\hline
	&\multicolumn{2}{c|}{\textbf{FFSIPP}} & \multicolumn{2}{c|}{\textbf{Baseline
			(SIPP)}} 	&\multicolumn{2}{c|}{\textbf{FFSIPP}} & \multicolumn{2}{c}{\textbf{Baseline
			(SIPP)}}
	\\		\cline{2-9}
	\textbf{SLA Level} & \textbf{Strict} & \textbf{Lenient} & \textbf{Strict}
	& \textbf{Lenient} & \textbf{Strict} & \textbf{Lenient} & \textbf{Strict}
	& \textbf{Lenient} \\ 	\cline{2-9}
	\hline \makecell[l]{\textbf{Number of}\\\textbf{Total Process Requests}}&
	\multicolumn{4}{c|}{50} & \multicolumn{4}{c}{100}
	\\
	\hline \makecell[l]{\textbf{Interval Between}\\\textbf{Process Requests}}
	& \multicolumn{4}{c|}{120 Seconds} & \multicolumn{4}{c}{60 Seconds}
	\\
	\hline \makecell[l]{\textbf{Number of Parallel}\\\textbf{Process Requests}} & \multicolumn{4}{c|}{5}  & \multicolumn{4}{c}{$f(n)$ -- see Equation~(\ref{eq:pyr})}
	\\\hline
	\makecell[l]{\textbf{SLA Adherence in \%} \\
		\textbf{(Standard Deviation)}}
	& \makecell[c]{98.67 \\ $\sigma = 1.15$}
	& \makecell[c]{100.00 \\ $\sigma = 0.00$}
	& \makecell[c]{86.00 \\ $\sigma = 2.00$}
	& \makecell[c]{98.67 \\ $\sigma = 1.15$}
	& \makecell[c]{97.67 \\ $\sigma = 0.58$}
	& \makecell[c]{100.00 \\ $\sigma = 0.00$}
	& \makecell[c]{95.33 \\ $\sigma = 0.58$}
	& \makecell[c]{99.67 \\ $\sigma = 0.58$}
	\\\hline
	\makecell[l]{\textbf{Total Makespan in Minutes} \\
	\textbf{(Standard Deviation)}}
	& \makecell[c]{27.33 \\ $\sigma = 1.15$}
	& \makecell[c]{36.67 \\ $\sigma = 2.08$}
	& \makecell[c]{29.33 \\ $\sigma = 0.58$}
	& \makecell[c]{29.33 \\ $\sigma = 0.58$}
	& \makecell[c]{63.33 \\ $\sigma = 0.58$}
	& \makecell[c]{69.67 \\ $\sigma = 2.89$}
	& \makecell[c]{62.00 \\ $\sigma = 0.00$}
	& \makecell[c]{61.67 \\ $\sigma = 0.58$}
	\\\hline
	\makecell[l]{\textbf{Leasing Cost} \\
		\textbf{(Standard Deviation)}}
	& \makecell[c]{1148.33 \\ $\sigma = 20.50$}
	& \makecell[c]{1102.00 \\ $\sigma = 14.42$}
	& \makecell[c]{2201.00 \\ $\sigma = 55.56$}
	& \makecell[c]{2052.67 \\ $\sigma = 147.99$}
	& \makecell[c]{2322.00 \\ $\sigma = 39.74$}
	& \makecell[c]{2181.67 \\ $\sigma = 89.44$}
	& \makecell[c]{4413.33 \\ $\sigma = 121.49$}
	& \makecell[c]{3975.00 \\ $\sigma = 118.87$}
	\\\hline
	\makecell[l]{\textbf{Penalty Cost} \\
		\textbf{(Standard Deviation)}}
	& \makecell[c]{1.00 \\ $\sigma = 1.00$}
	& \makecell[c]{0.00 \\ $\sigma = 0.00$}
	& \makecell[c]{15.00 \\ $\sigma = 2.65$}
	& \makecell[c]{0.67 \\ $\sigma = 0.58$}
	& \makecell[c]{3.33 \\ $\sigma = 1.53$}
	& \makecell[c]{0.00 \\ $\sigma = 0.00$}
	& \makecell[c]{10.67 \\ $\sigma = 1.15$}
	& \makecell[c]{0.33 \\ $\sigma = 0.58$}
	\\\hline
	\makecell[l]{\textbf{Total Cost} \\
		\textbf{(Standard Deviation)}}
	& \makecell[c]{1149.33 \\ $\sigma = 20.01$}
	& \makecell[c]{1102.00 \\ $\sigma = 14.42$}
	& \makecell[c]{2216.00 \\ $\sigma = 57.51$}
	& \makecell[c]{2053.33 \\ $\sigma = 148.37$}
	& \makecell[c]{2325.33 \\ $\sigma = 38.21$}
	& \makecell[c]{2181.67 \\ $\sigma = 89.44$}
	& \makecell[c]{4424.00 \\ $\sigma = 122.32$}
	& \makecell[c]{3975.33 \\ $\sigma = 119.43$}
	\\
\end{tabular}
\end{adjustbox}
\end{table*}

\paragraph{Pyramid Arrival of Resource-Intensive Processes} Next, we compare the enactment results for resource-intensive services for the pyramid arrival pattern as introduced in Equation~(\ref{eq:pyr}). An overview of the results is again given in Table~\ref{tab:evaluationReducedToBaseline:resourceIntensive:constant}.  Figure~\ref{subfig:pyramid15_intense:strict} shows the results for process requests with a strict deadline, while Figure~\ref{subfig:pyramid15_intense:lenient} depicts the results for lenient process requests.

Again, when compared to the baseline, our approach results in a better SLA adherence and lower associated penalty cost
as well as leasing cost that are approximately half as high. This time, the difference between the enactment of process requests
with a strict or a lenient deadline is relatively small. The explanation for the large saving of leasing cost of our approach is
the same as for the constant arrival pattern and can be attributed to the more fine-grained resource utilization which can be accomplished through FFSIPP.
Once again, our approach was also able to postpone steps into the future, leading to a longer total makespan for process requests
with lenient deadlines, when compared to the baseline.

It should be noted that lenient deadlines lead to better results in both our approach as well as the baseline, but an important
observation can be made when looking at the evaluation more closely. The performance benefit, especially in terms of cost is
much higher when considering lenient deadlines for the baseline approach. 

Our approach also yields a slightly better result for
lenient deadlines, but in comparison, this difference is rather small. For the baseline, a large number of requests that have to be
enacted in close temporal proximity result in having to lease individual
VMs for almost all service types. Some service types might only occupy a
small part of a leased machine while for other types multiple machines need to be leased.
More lenient  deadlines allow the use of already leased machines when computational resources are freed up in the near future, leading
to the observed performance benefit. Depending on the defined preferences and cost considering penalties, the system can accept
higher penalties for lower leasing cost. The higher the accepted penalties are, the more leasing cost could be saved.

In contrast to SIPP, our approach is able to use free resources from any leased VM to deploy a container of any
type with only as many assigned resources as the invocation of service requests requires. In contrast, the SIPP assumes that there is one service per VM running concurrently. This flexible of FFSIPP resource
utilization leads to comparable leasing cost for strict and lenient deadlines, as the number of leased but unused resources
is not increased as easily as for the baseline.

The results clearly show the main strength of FFSIPP, which is the better resource utilization due to
a fine-grained scheduling leading to overall lower leasing cost. We show this main strength while comparing our
approach to an already strong baseline, which delivers considerably better results in terms of cost and
SLA adherence than simple threshold-based ad-hoc scheduling strategies~\citep{hoenisch:16}.
The baseline approach already delivers a good SLA adherence as it detects potential violations in time and optimizes
the scheduling decision accordingly. Nevertheless, FFSIPP leads to even less SLA violations.

Both approaches allow SLA violations if the associated penalty fee leads to lower total cost than leasing additional resources
to finish process enactments in time. As our approach allows a more flexible scheduling of containers on VMs,
it can perform cheaper scheduling decisions and make better use of already leased resources, resulting in
lower SLA violations to leasing cost conflicts.

Generally, the results do not only depend on the number and type of process request arrivals and the available resources,
but also on the attributes of the process steps. In particular, service types with larger resource requirements will lead
to large containers taking up a great amount of the offered resources of a VM instance and, therefore, fewer resources
remaining available for other service requests of different service types. This consideration leads to the assumption
that our approach is capable of outperforming the baseline even more when service instances with smaller resource requirements
are considered. %We evaluate this assumption in the following section.

\subsubsection{Less Resource-intensive Services}
\label{subsub:lessresource}
\begin{table}[h]
	\caption{Evaluation Services -- Less Resource-Intensive}
	\label{tab:evalServicesLessResourceIntensive}
	\centering
		\footnotesize
	\begin{tabular}{c|c|c}
		\textbf{Service Type Name} & \textbf{CPU Load} &
		\textbf{Service Makespan} \\
		\textbf{} & \textbf{in \% ($\mu_{cpu}$)} & \textbf{in sec. ($\mu_{dur}$)} \\
		\hline A
		& 5
		& 40
		\\
		B
		& 10
		& 80
		\\
		C
		& 15
		& 120
		\\
		D
		& 30
		& 40
		\\
		E
		& 45
		& 100
		\\
		F
		& 55
		& 20
		\\
		G
		& 70
		& 40
		\\
		H
		& 125
		& 20
		\\
		I
		& 125
		& 60
		\\
		J
		& 190
		& 30
		\\
	\end{tabular}
\end{table}

\begin{figure}[t]
	\centering
	\subfloat[][Constant Arrival, Strict
	SLA]{\includegraphics[width=0.35\textwidth]{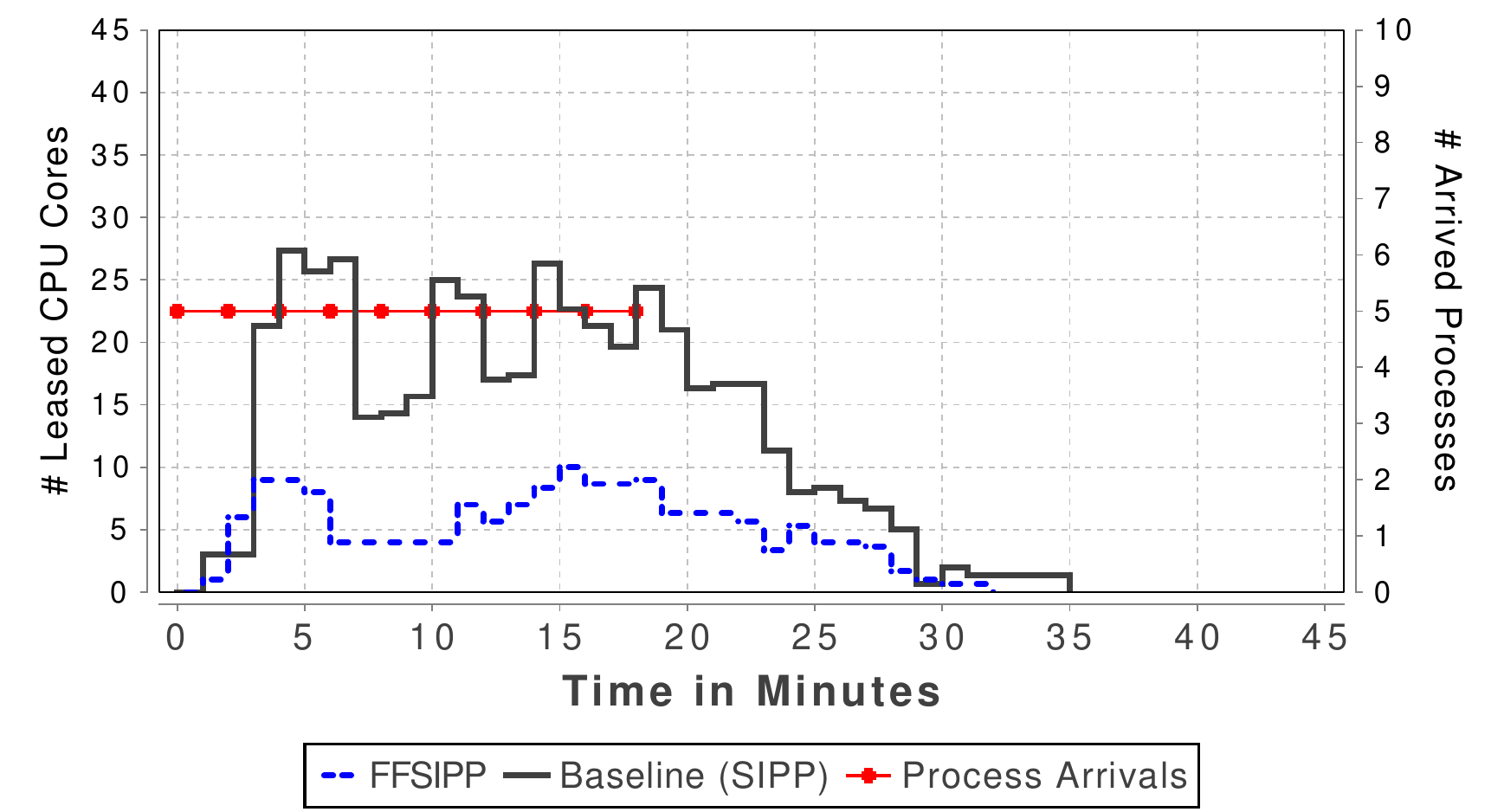}\label{subfig:constant15_lessIntense:strict}}
	\\
	\subfloat[][Constant Arrival, Lenient
	SLA]{\includegraphics[width=0.35\textwidth]{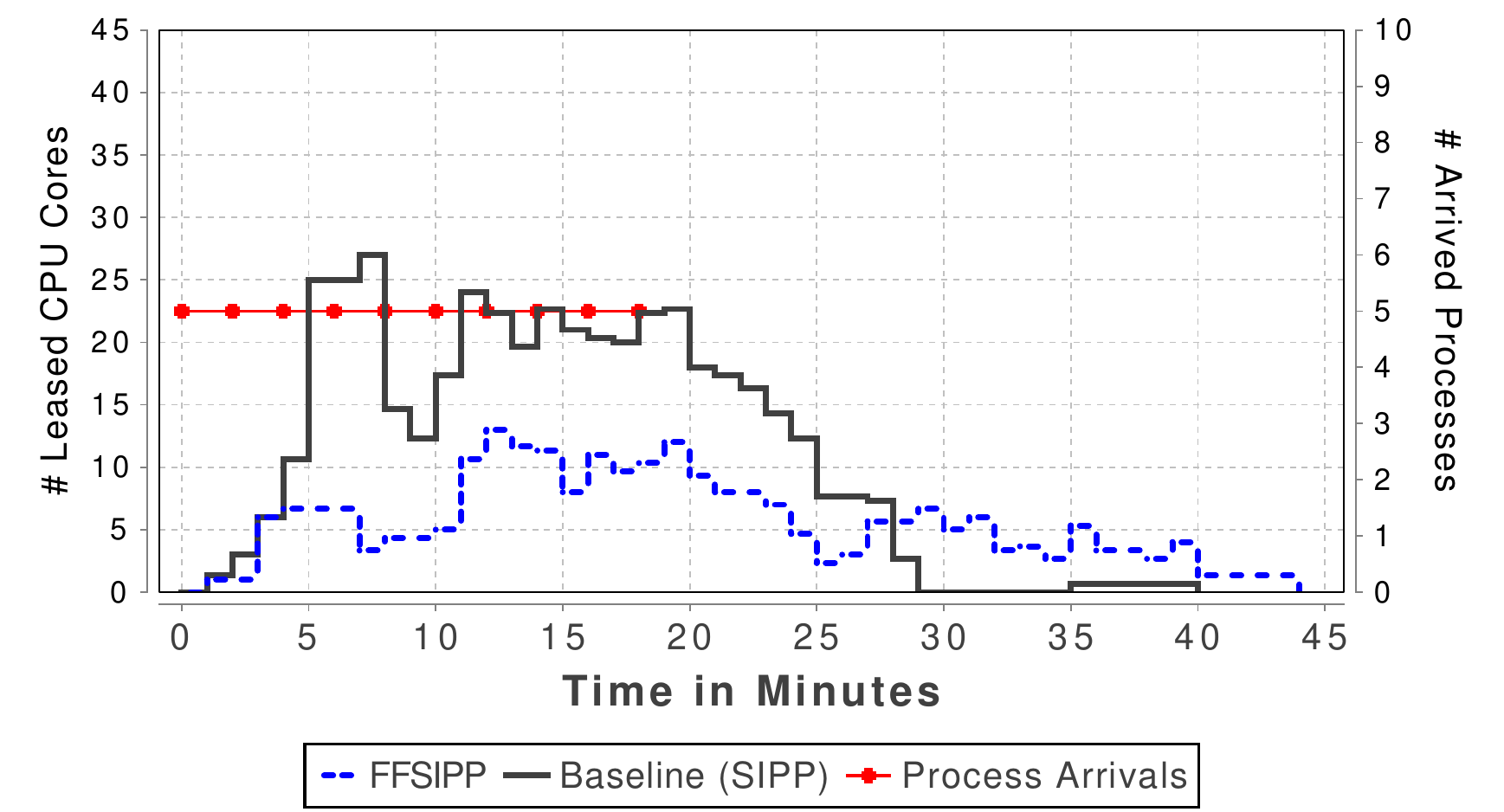}\label{subfig:constant15_lessIntense:lenient}}\\
	%\caption{Evaluation Results - Constant Arrival of Less Resource-Intensiv Processes}
	%	\label{fig:constant15_lessIntense}
	\caption{Evaluation Results -- Constant Arrival (Less Resource-Intensive Services)}
	\label{fig:constant}
\end{figure}

\begin{figure}[t]
	\centering
	\subfloat[][Pyramid Arrival, Strict
	SLA]{\includegraphics[width=.35\textwidth]{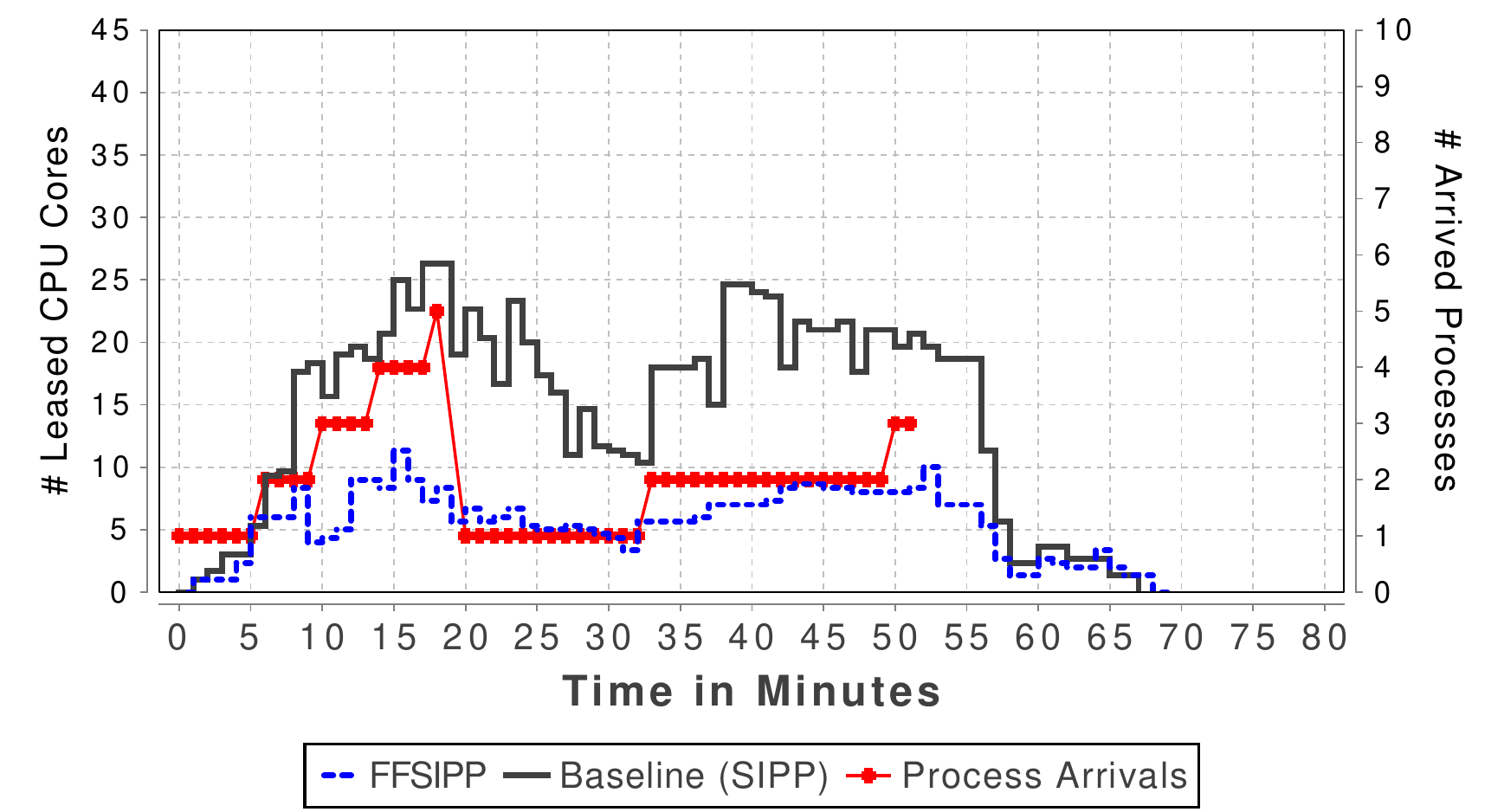}\label{subfig:pyramid15_lessIntense:strict}}
	\\
	\subfloat[][Pyramid Arrival, Lenient
	SLA]{\includegraphics[width=.35\textwidth]{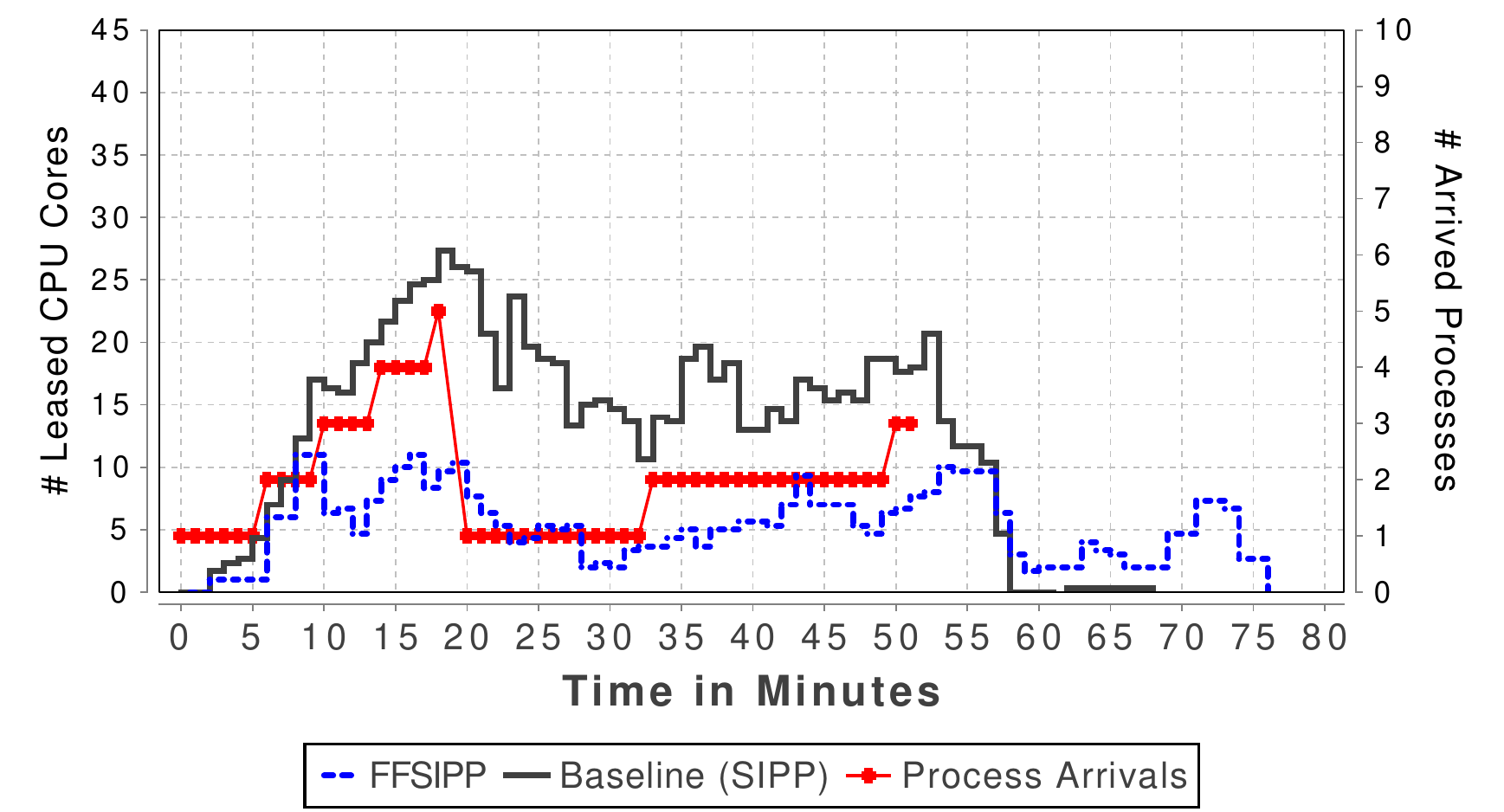}\label{subfig:pyramid15_lessIntense:lenient}}
	\caption{Evaluation Results -- Pyramid Arrival (Less Resource-Intensive Services)}
	\label{fig:pyramid}
\end{figure}

To evaluate this assumption, we perform the same evaluation runs as discussed before. However, we are now using the less resource-intensive service types introduced in Table~\ref{tab:evalServicesLessResourceIntensive}, instead of the previously used more resource-intensive service types from Table~\ref{tab:evalServicesResourceIntensive}.

\paragraph{Constant Arrival of Less Resource-Intensive Processes} The results for the constant arrival pattern and the less resource-intensive processes are presented in Table~\ref{tab:evaluationReducedToBaseline:lessResourceIntensive:constant}. 
Figure~\ref{subfig:constant15_lessIntense:strict} shows the results for process requests with a strict deadline, while
Figure~\ref{subfig:constant15_lessIntense:lenient} depicts the results for lenient process requests.

As expected, the results for process requests composed of less resource-intensive process steps show even higher
savings in terms of leasing cost when comparing our approach to the baseline, with around 2.5 times lower cost.
The SLA adherence of our approach for less resource-intensive processes is comparable to the evaluation of
the more resource-intensive processes as presented above, but the SIPP baseline shows
a higher SLA adherence with the less resource-intensive service types.

The improved SLA adherence for the baseline is easy to explain when considering that with smaller service types
more service invocations of the same type can be enacted on a VM instance, leading to a lower number of
penalty cost versus leasing cost conflicts for the baseline approach. As FFSIPP is able to place
requests on any already leased VM instance, there already exists a relatively large amount of scheduling
flexibility for the more resource-intensive service types, leading to similar SLA adherence values,
independent of the required resources for process steps.

All other observations already discussed for more resource-intensive process steps can also be observed in the evaluation runs discussed in this section, e.g., the consequences of more lenient deadlines.\\

\begin{table*}[t]
	\caption{Evaluation Results -- Less Resource-intensive
		Services}
	\label{tab:evaluationReducedToBaseline:lessResourceIntensive:constant}
	\centering

\begin{adjustbox}{width=0.85\linewidth,center}
	\begin{tabular}{l|c|c|c|c|c|c|c|c}
		\textbf{Arrival Pattern} &  \multicolumn{4}{c|}{\textbf{Constant Arrival}}
		& \multicolumn{4}{c}{\textbf{Pyramid Arrival}} 
		\\
		\hline
		&\multicolumn{2}{c|}{\textbf{FFSIPP}} & \multicolumn{2}{c|}{\textbf{Baseline
				(SIPP)}} 	&\multicolumn{2}{c|}{\textbf{FFSIPP}} & \multicolumn{2}{c}{\textbf{Baseline
				(SIPP)}}
		\\		\cline{2-9}
		\textbf{SLA Level} & \textbf{Strict} & \textbf{Lenient} & \textbf{Strict}
		& \textbf{Lenient} & \textbf{Strict} & \textbf{Lenient} & \textbf{Strict}
		& \textbf{Lenient} \\ 	\cline{2-9}
		\hline \makecell[l]{\textbf{Number of}\\\textbf{Total Process Requests}}&
		\multicolumn{4}{c|}{50} & \multicolumn{4}{c}{100}
		\\
		\hline \makecell[l]{\textbf{Interval Between}\\\textbf{Process Requests}}
		& \multicolumn{4}{c|}{120 Seconds} & \multicolumn{4}{c}{60 Seconds}
		\\
		\hline \makecell[l]{\textbf{Number of Parallel}\\\textbf{Process Requests}} & \multicolumn{4}{c|}{5}  & \multicolumn{4}{c}{$f(n)$  -- see Equation~(\ref{eq:pyr})}
		\\
		\hline \makecell[l]{\textbf{SLA Adherence in \%} \\
			\textbf{(Standard Deviation)}}
		& \makecell[c]{97.33 \\ $\sigma = 1.15$}
		& \makecell[c]{99.33 \\ $\sigma = 1.15$}
		& \makecell[c]{93.33 \\ $\sigma = 2.31$}
		& \makecell[c]{99.33 \\ $\sigma = 1.15$}
		& \makecell[c]{97.33 \\ $\sigma = 0.58$}
		& \makecell[c]{100.00 \\ $\sigma = 0.00$}
		& \makecell[c]{98.67 \\ $\sigma = 0.58$}
		& \makecell[c]{100.00 \\ $\sigma = 0.00$}
		\\
		\hline \makecell[l]{\textbf{Total Makespan in Minutes} \\
			\textbf{(Standard Deviation)}}
		& \makecell[c]{27.33 \\ $\sigma = 1.15$}
		& \makecell[c]{37.33 \\ $\sigma = 3.06$}
		& \makecell[c]{29.67 \\ $\sigma = 1.15$}
		& \makecell[c]{31.00 \\ $\sigma = 4.36$}
		& \makecell[c]{63.67 \\ $\sigma = 2.31$}
		& \makecell[c]{71.67 \\ $\sigma = 0.58$}
		& \makecell[c]{61.67 \\ $\sigma = 1.15$}
		& \makecell[c]{63.33 \\ $\sigma = 4.04$}
		\\
		\hline \makecell[l]{\textbf{Leasing Cost} \\
			\textbf{(Standard Deviation)}}
		& \makecell[c]{479.33 \\ $\sigma = 36.07$}
		& \makecell[c]{512.00 \\ $\sigma = 32.14$}
		& \makecell[c]{1316.67 \\ $\sigma = 124.16$}
		& \makecell[c]{1283.67 \\ $\sigma = 187.79$}
		& \makecell[c]{1018.67 \\ $\sigma = 25.97$}
		& \makecell[c]{996.67 \\ $\sigma = 48.33$}
		& \makecell[c]{2619.00 \\ $\sigma = 91.02$}
		& \makecell[c]{2430.33 \\ $\sigma = 114.29$}
		\\
		\hline \makecell[l]{\textbf{Penalty Cost} \\
			\textbf{(Standard Deviation)}}
		& \makecell[c]{2.33 \\ $\sigma = 0.58$}
		& \makecell[c]{1.33 \\ $\sigma = 2.31$}
		& \makecell[c]{6.67 \\ $\sigma = 1.15$}
		& \makecell[c]{0.33 \\ $\sigma = 0.58$}
		& \makecell[c]{4.33 \\ $\sigma = 1.15$}
		& \makecell[c]{0.00 \\ $\sigma = 0.00$}
		& \makecell[c]{3.6 \\ $\sigma = 0.58$}
		& \makecell[c]{0.00 \\ $\sigma = 0.00$}
		\\
		\hline \makecell[l]{\textbf{Total Cost} \\
			\textbf{(Standard Deviation)}}
		& \makecell[c]{481.67 \\ $\sigma = 36.12$}
		& \makecell[c]{513.33 \\ $\sigma = 34.44$}
		& \makecell[c]{1323.33 \\ $\sigma = 125.13$}
		& \makecell[c]{1284.0 \\ $\sigma = 188.34$}
		& \makecell[c]{1023.00 \\ $\sigma = 24.88$}
		& \makecell[c]{996.67 \\ $\sigma = 48.34$}
		& \makecell[c]{2622.67 \\ $\sigma = 90.47$}
		& \makecell[c]{2430.33 \\ $\sigma = 114.29$}
		\\
	\end{tabular}
\end{adjustbox}
\end{table*}

\paragraph{Pyramid Arrival of Less Resource-Intensive Processes} Finally, we compare the results for less resource-intensive processes for the pyramid arrival pattern as introduced in Equation~(\ref{eq:pyr}).
Table~\ref{tab:evaluationReducedToBaseline:lessResourceIntensive:constant} again gives an overview of the results. Figure~\ref{subfig:pyramid15_lessIntense:strict} shows the results for process requests with a strict deadline while
Figure~\ref{subfig:pyramid15_lessIntense:lenient} depicts the results for lenient process requests.

Again, we can observe that our approach leads to about 2.5 times lower leasing cost for process requests composed of
less resource-intensive process steps when compared to the baseline. We can also make the same observations for SLA
adherences as we have already done for the constant arrival of less resource-intensive processes, as well as the other observations
discussed above for more resource-intensive process steps. However, there is one notable exception: For the strict deadlines, the baseline provides a better SLA adherence of 98.67\% compared to FFSIPP's 97.33\%. However, this comes at the price of a cost increase of 156.37\%.

\subsection{Summary and Discussion}
In this section, we presented an extensive evaluation of the FFSIPP approach. In the evaluation, we considered a number of evaluation process models, different process request arrival patterns and system settings, and presented the associated enactment results. We evaluated the results of the MILP-based FFSIPP combined with the transformation step, comparing it to a state-of-the-art approach which does not take into account containers. We showed that due to the overall more
efficient use of leased computational resources, our approach calculates scheduling results with 2 to 2.5 times lower leasing cost if compared to the baseline. FFSIPP also outperforms the baseline with regard to SLA adherence, achieving adherence levels of at least 97\% for all evaluated scenarios.

Despite the promising results, it should be taken into account that the evaluation could of course be further extended. For instance, the evaluation setting is relatively small, with 50 and 100 process requests, respectively. As mentioned in Section~\ref{sub:transformation}, MILP-based optimization becomes difficult to achieve if the problem sizes increase. The reason for this is that the scheduling of services is an NP-hard problem~\cite{hoenisch:16}, i.e., it is unknown if a solution to the optimization problem can be found in polynomial time. Given the large number of variables, the optimization problem shows scalability issues if too many process requests have to be handled at the same time. This could be solved by applying a heuristic algorithm to solve the optimization problem, e.g., genetic algorithms~\cite{WYSM18}.

\section{Related Work}
\label{sec:related}
Elastic processes have gained quite a lot of attention by the cloud computing and BPM research communities in recent years. In the following paragraphs, we discuss relevant concepts for optimized scheduling and resource allocation for elastic processes. 

To start with, a multitude of solutions for task scheduling and resource allocation have been proposed for single services and applications and aiming at the VM level, e.g.,~\citep{wu:11, van:13}. In addition, some approaches regard resource allocation and task scheduling on the container level: \cite{xu:14} propose a game-theoretic model, aiming to find a solution which reduces the response times while improving the resource utilization of cloud providers. \cite{hoenisch:15} formulate a multi-objective optimization problem for finding an optimal resource allocation and task scheduling. Importantly, the presented solution aims at both the VM level and the container level, i.e., the authors combine horizontal and vertical scaling, as it is also the foundation for the work at hand. \cite{nardelli:17} propose an Integer Linear Programming problem that considers heterogeneous container requirements and VM resources when calculating an optimal allocation as well as a runtime reallocation of containers on a set of leased VMs. %More recently, \cite{dinitto20} have presented \emph{Gru}, which provides a holistic stack to self-management for microservices. The authors discuss a number of policies and operational strategies to realize an autonomic microservice cluster.

The discussed approaches focus on cost measures based on the actual utilization of cloud-based computational resources as well as breaches of negotiated SLAs, but generally do not consider the process perspective. %, such as the relations and dependencies among multiple process steps and the underlying services. 
Therefore, findings from resource allocation and task scheduling for single services can not be directly mapped to elastic processes. %Nevertheless, these approaches provide valuable insights that can be used as a foundation for the optimization of elastic process enactments. 

\cite{herrera20} propose an auto-scaling mechanism for the orchestration of containers. For this, the authors present a number of bio-inspired algorithms which are used to scale the computational resources. \cite{casalicchio19} presents a survey and a reference architecture for container orchestration. %Here, the authors state that the number of actual solutions for container orchestration is still quite small. 
It should be noted that container orchestration is usually not based on complex process patterns as we have applied in the work at hand.

For the realization of elastic processes, it is especially necessary to account for the data and control flows that come with the enactment of processes~\citep{schulte:15}. Again, a number of resource allocation and task scheduling solutions which aim at VMs (instead of containers) have already been proposed: \cite{xu:09} present a solution for scheduling multiple workflows with QoS constraints in the cloud. The proposed scheduling algorithm considers factors that affect the total makespan and cost of workflows, and aims at improving the mean enactment time and enactment cost of all workflows in a process landscape, while taking into account QoS constraints. A multi-objective genetic scheduling algorithm for BPEL workflows in distributed cloud environments is presented by \cite{juhnke:11}. Their approach considers data dependencies between BPEL workflow steps and uses workflow enactment times as well as cost for the needed computational resources. By using weights, the Pareto-optimal solution of the multi-objective heuristic can be transformed into a unique solution. A number of strategies that are based on similar assumptions have also been proposed by \cite{bessai:13}. The authors aim at scheduling process steps of business processes using elastic cloud-based resources, while optimizing enactment time and cost. Both the approaches of \cite{juhnke:11} and \cite{bessai:13} provide solutions that allow for a parallel enactment of processes, but in contrast to our work, do not account for SLAs. 

SLAs like deadlines for single tasks and processes have been considered for sequential processes by \cite{wei:16}. The authors introduce an adaptive configuration algorithm for dynamically managing VMs for service-based workflows in the cloud.
%Services are assigned to VM instances, which are in turn aggregated on physical machines. 

%ViePEP, which is used as the eBPMS in the work at hand, has originally been developed for usage with VMs~\citep{hoenisch:16, hoenisch:15:iccc}. In our former work, we also present resource allocation and task scheduling approaches for elastic processes. Similar to the work at hand, we used MILP for the optimization, aiming at minimizing cost while taking into account process deadlines. 

\cite{rosinosky:17, rosinosky18} discuss an ILP-based solution and a genetic algorithm that aim at finding an efficient resource allocation for elastic processes while reducing the number of migrations from one VM to another VM. In contrast to the work at hand, processes are not divided into steps. Hence, in our work, we aim at providing computational resources on a more fine-grained level. 

None of the so-far mentioned solutions makes use of containers. Hence, in the work at hand, we propose an approach to exploit the advantages of container-based virtualization approaches, allowing a more fine-grained model for resource allocation and task scheduling for elastic processes. To the best of our knowledge, the related work explicitly aiming at the optimization on the container-level is still very limited. 

Most existing approaches have been proposed in the area of scientific workflows (SWFs), e.g.,~\citep{zheng:17, lopez19}. The main limitation of these approaches is that they do not consider concurrent processes, which are the norm in the area of BPM. Furthermore, SWFs are generally more data flow-oriented, while business process scheduling is used in complex landscapes with concurrent process requests. Business processes often share the same services and tend to be more control flow-driven than SWFs~\citep{ludascher:09}. Hence, solutions from the field of SWFs give some interesting ideas, but cannot be directly applied to BPM. 

In our former work~\citep{WHS+19,WYSM18}, we discuss an updated version of ViePEP, which is able to use containers instead of VMs for service deployment. We also extend our optimization models for resource allocation and task scheduling by allowing optimization both on the VM and on the container level. For this, we apply genetic algorithms, but have so far not provided an  optimal solution on the VM level as we do in the work at hand by formulating a MILP problem.

The idea of using microservices for realizing business process steps is discussed by \cite{andrews18}. The authors make use of a distributed, actor-based approach, where actors are connected to each other. This allows high scalability, but makes it necessary that the actors are aware of each other and of the underlying process models. 

\cite{boukadi17,boukadi19} also make use of containers for process enactment. Similar to the work at hand, the goal is to find a cost-optimal solution to the resource allocation and task scheduling problem. Notably, the authors allow scaling on the level of VMs and a subsequent allocation of containers, i.e., also combine horizontal and vertical scaling. For this, the authors apply linear programming and genetic algorithms. However, optimization is only done for single processes instead of complete process landscapes, and containers (respectively the underlying computational resources) cannot be shared between processes. Process deadlines are not regarded. In contrast to the work at hand, \cite{boukadi17,boukadi19} do not allow a trade-off between SLA breaches and cost through the integration of penalty fees. %Also, they make use of simulations instead of a real-world testbed. 
Nevertheless, the approach by \cite{boukadi17,boukadi19} comes closest to our work presented here.

Last but not least, it should be noted that our work makes use of worst-case assumptions for chosen paths in a process model, startup times of containers, etc. While this is a very common approach in business process and service research, it could of course also be interesting to take into account some predictions about process behavior in order to avoid the complex and conservative worst-case analyses applied in our work. Most importantly, predicting the next process step, as proposed by \cite{evermann17} for implicitly defined process models, is necessary for the work at hand. Going one step further by incorporating approaches to predict the chosen process path, e.g.,~\citep{polato18}, would allow us to decrease the search space of the optimization, which could result in faster results which fit the actual process enactments even better.

%To conclude the discussion of the related work, it should be highlighted that most related approaches make use of VMs instead of containers as the underlying computational resources to host services. Notably, to the best of our knowledge, there is no other solution which aims at four-fold optimization of resource allocation and task scheduling, i.e., combines horizontal and vertical scaling.

%NOTE: If we need to fill additional space, we could also discuss SWFs, Gerta has done this actually very fine at the beginning of Section 3.2.1
\section{Conclusion}
\label{sec:conclusion}
Elastic processes, i.e., business processes enacted on elastic cloud infrastructure, have so far been primarily regarded on a rather coarse-grained level, with VMs being used as computational resources. The utilization of container technologies in order to provide more fine-grained control over cloud-based computational resources has the potential to save cost and to benefit from startup times that are significantly smaller. % has not gained a lot of attention so far. 
One particular research question in this area is how to make sure that the process steps are carried out in a cost-efficient way while taking into account the QoS requirements of the process owners.

In this paper, we introduced a multidimensional optimization approach, which is combined with a transformation step in order to realize cost-efficient auto-scaling of container-based elastic processes. Our approach enables fine-grained resource allocation and task scheduling for elastic processes. Scaling is achieved over four dimensions, i.e., horizontally and vertically, both for VM resources and Docker containers. The approach focuses on the minimization of VM leasing cost as well as penalty cost for SLA violations. % while considering times for VM and container leasing, startup, and deployment.
As the evaluation has shown, this leads to significant cost reductions in the enactment of business process landscapes, up to a factor of 2 to 2.5 compared to a state-of-the-art approach. Also, our solution further improves the already very good SLA adherence of the state-of-the-art.

In our future work, %we want to further extend our approach by taking into account prediction-based process patterns instead of the current worst-case analysis. %This allows to explicitly regard the probability that, e.g., a particular XOR-path is enacted, and therefore decreases the complexity of the optimization problem to some degree. 
we want to integrate approaches which predict the process request arrival patterns based on historical data. This way, the FFSIPP could be enhanced from a reactive into a predictive approach. Also, we want to take into account that process makespans may also include communication and data transfer overhead, by extending our system and optimization model accordingly. Last but not least, applying serverless approaches could lead to an even higher level of flexibility, and is therefore an interesting option for the future.

\section*{Acknowledgements}
The financial support by the Austrian Federal Ministry for Digital and Economic Affairs, the National Foundation for Research, Technology and Development as well as the Christian Doppler Research Association for the Christian Doppler Laboratory for Blockchain Technologies for the Internet of Things is gratefully acknowledged.

%\ifCLASSOPTIONcompsoc
%\else
%\fi
%This work is partially supported and funded by the Austrian Research Promotion Agency (FFG) via the ``Austrian Competence Center for Digital Production'' (CDP) under the contract number 854187.
%The majority of the work was done while the first and the fourth author were with Data61, CSIRO, in Sydney, Australia.

%\textbf{NOTE: Regular papers have 12 pages, max. 6 additional pages can be bought for \$ 220 each}

\bibliographystyle{elsarticle-num} 
\bibliography{bibliography}

\end{document}